\documentclass[12pt]{article}
\usepackage{graphicx}
\usepackage{amssymb}

\begin{document}
\begin{center}
{\Large\textbf{Quasicrystals---The impact of N.G.~de Bruijn}}

\vspace{1em}
Helen Au-Yang and Jacques H.H.~Perk
\vspace{1em}

Department of Physics, Oklahoma State University,\\
145 Physical Sciences, Stillwater, OK 74078-3072, USA
\end{center}
\begin{abstract}
In this paper we put the work of Professor N.G.~de Bruijn on\break
quasicrystals in historical context. After briefly discussing what
went before, we shall review de Bruijn's work together with
recent\break related theoretical and experimental developments. We conclude
with\break a discussion of Yang--Baxter integrable models on Penrose tilings,
for which essential use of de Bruijn's work has been made.
\end{abstract}


\section{Introduction\label{sec1}}

\subsection{Dedication}

This article is devoted to the memory of one of the great mathematicians
of the twentieth century, Prof.\ dr.\ N.G.\ de Bruijn. For this purpose,
we shall describe here the important and unique contributions of de Bruijn
to the development of the theory of quasicrystals together with their
historical context.

We shall review several theoretical developments related to de Bruijn's
work. We shall also briefly discuss Shechtman's experimental findings
for which he was awarded the 2011 Nobel Prize in Chemistry and mention
some recent technological applications that resulted from his discovery.
Finally, we shall discuss some of our own results, explaining explicitly
how de Bruijn's work has been used, but without going into the full
mathematical details.


\subsection{Pentagrid and Cut-and-project}

Two of de Bruijn's papers \cite{Bruijn1,Bruijn2} give deep
new insight into the nature of Penrose tilings \cite{Penrose,Pen1,Gardner}
introduced by Penrose in 1974 as a mathematical curiosity and game.
Penrose introduced two kinds of tiles, together with local matching
rules forcing a tiling with a five-fold rotational symmetry axis.

Rather than constructing the tiling using local matching rules for the
tiles that force aperiodicity, de Bruijn discovered two global construction
methods starting with a five-dimensional lattice. The first method,
a cut-and-project method, draws a two-dimensional plane and projects
all lattice points in the five-dimensional lattice to that plane, that are
within a certain distance from the plane as defined through a ``window."
The other method introduces a ``pentagrid" of five sets (grids) of
equidistant parallel lines, with lines of different grids intersecting
each other with angles of $36^{\circ}$ and $72^{\circ}$; the Penrose
rhombus tiling follows then by dualization, choosing each rhombus
according to the orientation of the two grid lines meeting at a vertex.

The pentagrid of de Bruijn is intimately related with the Conway worms
mentioned already in Gardner's 1977 article \cite{Gardner}. It is less directly
related with the Ammann bars \cite{Gardner2,GrSh}---special non-equidistant
lines, which have the advantage, however, that one can draw them on the
Penrose tiling itself, while they also define the tiling.
Therefore, it was not immediately clear to many physicists how great
de Bruijn's contribution is. His is a simpler, complete, systematic
and precise, mathematical description, allowing explicit calculations. The
findings of Conway and Ammann, as reported, e.g., by \cite{Gardner2,GrSh},
were mostly left as observations with details not worked out.
The advantage of the approach of de Bruijn is particularly clear in our
work discussed in the final section \ref{PIM}.

Quasiperiodic structures had been studied by mathematicians and even
appeared decorating medieval Islamic buildings. But because of the
two constructions of de Bruijn we may feel justified to call the
Penrose tiling a quasicrystal.


\subsection{Nobel Prize in Chemistry}

In 1982, while at the U.S.\ National Bureau of Standards near Washington,
DC,\ Dan Shechtman observed a ten-fold scattering pattern from a
metal alloy sample, that looked like an ordinary crystal. As this was
contradicting long-held beliefs in the physics and chemistry communities,
Shechtman was opposed by many colleagues---and even ridiculed by some of
them---since his observations contradicted standard physics texts like
Kittel's \cite{Kittel}. Although initially rejected by referees,
Shechtman succeeded in publishing his findings in 1984 \cite{SBGC,SB}.

The early mathematical results on Penrose tilings and three-dimensional
structures as discussed
by Steinhardt et al.\ \cite{LSt0,LSt,SoSt} helped convince many
of Shechtman's colleagues, especially since Mackay
had even calculated a scattering pattern with ten-fold symmetry from
a Penrose tiling \cite{Mackay1,Mackay2}. In the end, few continued their
opposition, with the most prominent one being Nobel Prize winner
Linus Pauling. More examples of quasicrystals were found in the
laboratory and quasicrystals found several technical applications.
For his pioneering work Shechtman received the 2011 Nobel Prize in
Chemistry \cite{Nobel} for the discovery of quasicrystals.

In his Nobel lecture \cite{Nobel2} Shechtman compared himself with
a little pussycat trotting in front of a row of German shepherds, while
quoting Psalm 23 in the original Hebrew and King James English,
``Yea, though I walk through the valley of the shadow of death,
I will fear no evil." He also quoted from Kittel's authoritative
and widely-used textbook,
``We cannot find a lattice that goes into itself under other rotations,
such as by $2\pi/5$ radians or $2\pi/7$ radians. $\ldots$ We can make a
crystal from molecules which individually have a five-fold rotation
axis, but we should not expect the lattice to have a five-fold
rotation axis." \cite[p.~12]{Kittel}. These two passages in the Nobel
lecture illustrate how hard it was for Shechtman to get his findings
accepted by his colleagues. Finally, in the seventh edition of his
textbook Kittel acknowledged the existence of quasicrystals
\cite[pp.\ 48, 49]{Kittel2}.


\subsection{Applications of Quasicrystals}

Quasicrystals exhibit very unusual properties and because
of these they have found many technological applications
\cite{Lutz,Jacoby}. The 2011 Nobel Committee for Chemistry wrote
\cite{LTF}: ``When trying out different blends of metal,
a Swedish company managed to create steel with many surprisingly
good characteristics. Analyses of its atomic structure showed
that it consists of two different phases: hard steel quasicrystals
embedded in a softer kind of steel. The quasicrystals function
as a kind of armor." Quasicrystals are also bad conductors
of heat and electricity and have non-stick surfaces.

Therefore, the number of applications of quasicrystals is ever increasing.
Quasicrystals are now used in razor blades, thin needles made for eye
surgery, materials for reuse of waste heat, surface coatings for
frying pans (causing the pan to heat evenly), energy-saving light-emitting
diodes (LEDs), heat insulation in engines (especially diesel engines),
see e.g.\ \cite{Lutz,Jacoby}.

It may be noted that quasiperiodicity also can be generated in
optical lattices \cite{GDdSV}, which may lead to a different kind
of applications.


\section{Prehistory} 

\subsection{Islamic Art}

One of the earliest forerunners of quasiperiodic tilings is manifested in
the decoration of certain Islamic buildings \cite{Bohannon,LuSt,Ajlouni}.
Some examples are the Alhambra Palace, Spain, the Darb-i Imam Shrine,
Isfahan, Iran, and a Madrasa, in Bukhara, Uzbekistan. As the Muslim faith
did not allow images of life objects, artists and architects in Islamic
countries were searching for a wider variety of geometrical structures
to decorate their important buildings.


\subsection{D\"urer's Unterweysung der Messung}

Albrecht D\"urer (1471--1528) published a systematic study of
tilings in his ``Unterweysung der Messung" (1525). He concluded that
regular triangles, squares and hexagons can fill the plane, but
pentagons cannot by themselves do that without rhombi to fill
the holes, see \cite[bk.~2, 4~figs.~24]{durer}.

Similarly, twelve pentagons make a dodecahedron, twenty equilateral
triangles form an icosahedron (the dual of a dodecahedron). However,
these objects cannot be stacked without either leaving free space
{\it or} allowing partial overlaps.


\subsection{Fibonacci sequences}

Leonardo Fibonacci of Pisa ($\pm$1170 -- $\pm$1250) not only
advocated the use of Arabic notation of numbers over the Roman
numerals, he also is the originator of the Fibonacci sequence,
\begin{equation}
1,\,1,\,2,\,3,\,5,\,8,\,13,\,21,\,34,\,55,\ldots,
\qquad F_{n+1}=F_n+F_{n-1}.
\end{equation}
This describes the exponential growth of a rabbit population: Every
time step each pair produces a new pair that needs one time interval
to mature, before having their own offspring.

It is well-known that the Fibonacci numbers $F_n$ can be expressed
in terms of the golden ratio, which is commonly denoted by either
$\tau$ or $\phi$. Explicitly,
\begin{equation}
F_n={\tau^n-(1-\tau)^n\over\tau-(1-\tau)},\qquad
\lim_{n\to\infty}{F_{n+1}\over F_n}=\tau,
\end{equation}
where
\begin{equation}
\tau={1+\sqrt{5}\over2}=1.61803\cdots,\qquad
1-\tau=-{1\over\tau}={1-\sqrt{5}\over2}=-0.61803\cdots.
\end{equation}
These are the solutions of $x^2=x+1$.

Much more recent is the geometric Fibonacci sequence:
Assume $A$ is a red line piece of length 1 and $B$ a blue line
piece of length $\tau$. We can then concatenate them (link them together)
using the rule,
\begin{equation}
F_1=A,\quad F_2=B,\quad F_3=BA,\quad F_4=BAB,\quad F_5=BABBA,
\end{equation}
\begin{equation}
F_6=BABBABAB\,\ldots,\quad
F_{n+1}=F_nF_{n-1}.
\end{equation}
The resulting $F_{\infty}$ is quasiperiodic, i.e.\ a one-dimensional
quasicrystal. A generalization of this sequence has been studied
also by de Bruijn \cite{Bruijn0,Bruijn6}.


\subsection{Kronecker's Theorem\label{kron}}

Kronecker formulated a well-known theorem that implies quasiperiodic
sequences, known as Kronecker's Approximation Theorem \cite{HW}:
If $x$ is an irrational number, then the infinite sequence
$\{x\},\{2x\},\{3x\},\ldots$, is uniformly and
densely distributed within the unit interval.
Here $\{ nx\}\equiv nx-\lfloor nx\rfloor$ is the
fractional part of $nx$, i.e. $0\le\{ nx\}<1$.
Also, the formulation of the theorem given here is a more or
less obvious extension of the original \cite{HW}.


\subsection{Harald Bohr---Almost periodic functions}

Harald Bohr was a star of the Danish national soccer team,
that won the silver medal in the 1908 Olympics beating France by 17-1
(the all-time Olympic record). During Bohr's 1910 PhD defense the
room was filled with soccer fans. How unusual!

In his work on almost periodic functions,
Bohr gave several constructions of functions that are not periodic,
but almost repeat, i.e.\ $|f(t+T)-f(t)|<\epsilon$, for some
$T(\epsilon)$ for given arbitrarily small $\epsilon$ \cite{Bohr}.
This is a highly nontrivial generalization of Kronecker's sequence.


\subsection{Optical example---Fraunhofer diffraction pattern}

One can make the following variation of Young's double slit experiment:
\begin{figure}[htbp]
\begin{center}
\includegraphics[height=7cm]{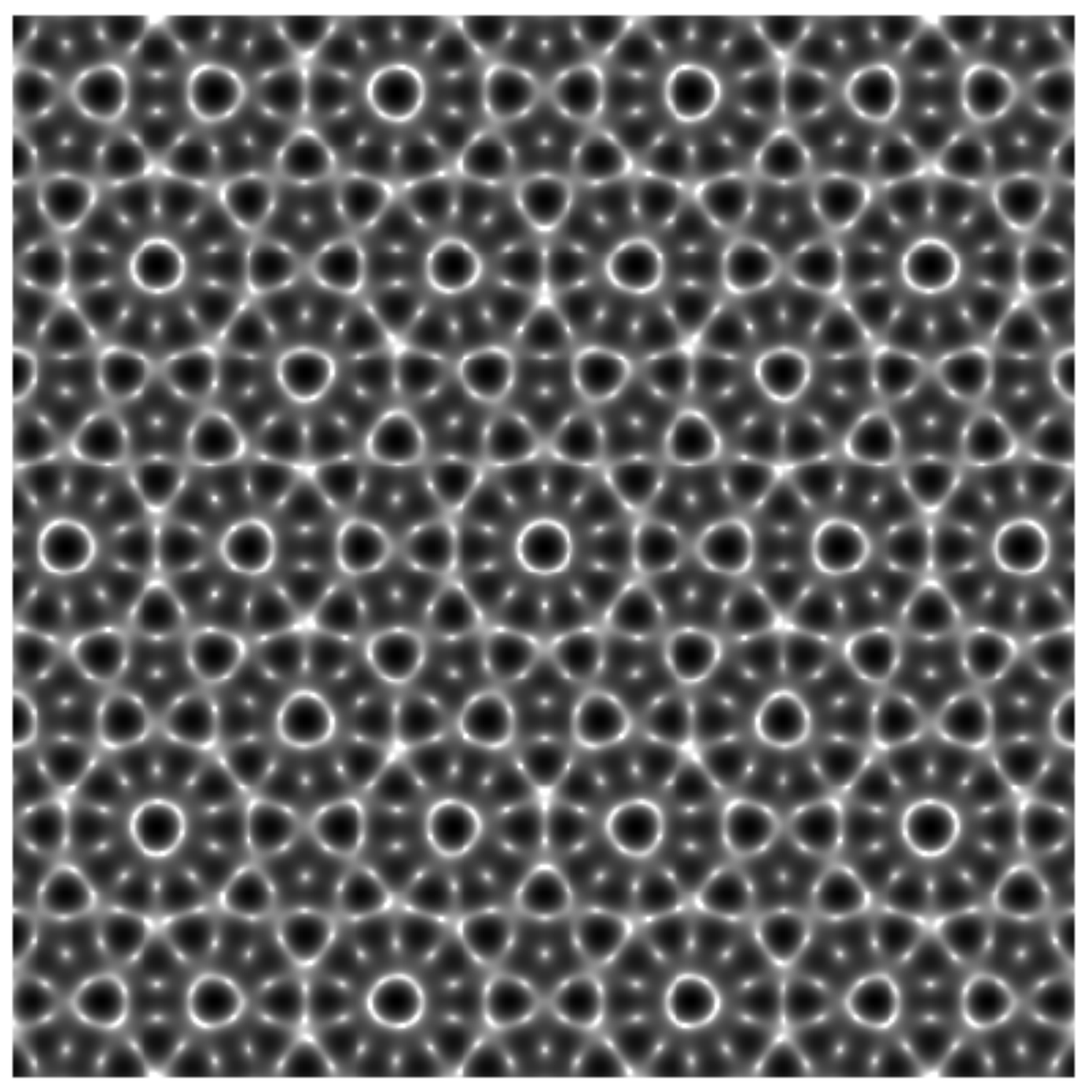}
\end{center}
\caption{Fraunhofer diffraction pattern from five pinholes arranged as
a pentagon. The darker the spot, the higher the intensity. Note that the
center area is almost repeated at ten places near the border.}
\label{fig1}\end{figure}
Illuminate five pinholes arranged as a pentagon in a screen
and observe the diffraction pattern on a far-away screen. Then
the local intensity on that screen is given by $I=I_0|A|^2$, with
\begin{equation}
\displaystyle{A(x,y)=\sum_{j=1}^5 {\mathrm e}^{{\mathrm i}
x\cos({{\scriptstyle2\pi}\over\scriptstyle5}j)
+{\mathrm i}y\sin({{\scriptstyle2\pi}\over\scriptstyle5}j)}}
\end{equation}
an almost-periodic analytic function of $x$ and $y$.

It is very easy to plot the intensity pattern using Maple,
for example:
\begin{verbatim}
f:=(x,y)->-log(abs(sum(exp(I*(x*cos(2*Pi*j/5)
   +y*sin(2*Pi*j/5))),j=0..4))^2);
with(plots): d:=50; g:=400:
densityplot(f(x,y),x=-d..d,y=-d..d,grid=[g,g],
   style=PATCHNOGRID,axes=NONE,scaling=CONSTRAINED);
\end{verbatim}
The resulting pattern, see figure \ref{fig1}, has ten-fold rotational symmetry and
thus cannot be periodic.


\section{Penrose Tilings}

Sir Roger Penrose, born in 1931, is a British mathematician with major
contributions to general relativity and cosmology. He introduced
Twistor Theory, mapping (3+1)-dimensional space-time and objects
therein to a space with 2-2 signature. He also contributed to the
theories of Black Holes and the Big Bang. He made some unusual
observations. For example, he calculated the entropy of the universe
at the Big Bang and found it to be extremely small, estimating
the ``Creator's aim" to be 1 in $10^{10^{123}}\!$
\cite[p.\ 344]{Pen9}.

Penrose discovered his tilings in the early 1970s and patented them
as they could be used for interesting puzzles and games. He published
his invention in 1974 \cite{Penrose,Pen1} and few people knew about
them before the 1977 paper of Martin Gardner \cite{Gardner,Gardner2}.

From the pentagon you can make two prototiles in different ways,
one way being a skinny rhombus ($36^{\circ}$-$144^{\circ}$) and 
a fat rhombus ($72^{\circ}$-$108^{\circ}$). Next Penrose introduced
matching rules to force the tiles into a quasiperiodic tiling:
The markings on neighboring tiles have to exactly match. One can use
different arrows on the edges, or colored arcs in the bodies, or
matching edge deformations (to make the jigsaw effect).
One can find similar matching rules for other pairs of prototiles
also, like Penrose Kites and Darts \cite{Penrose,Pen1}. The result
is an aperiodic tiling with the center being a five-fold rotation axis.

A well-known related tiling is the Ammann--Beenker tiling using
squares and $45^\circ$-$135^\circ$ rhombi. In this case the center is
an eight-fold rotation axis.

It is OK to stand on a Penrose tiling like the one at Texas A \&\ M,
but when Kleenex sold toilet paper with a Penrose tiling design on it,
Penrose sued the company saying that his patent was violated.


\section{Nicolaas Govert de Bruijn (1918-2012)}

N.G.\ de Bruijn started his work on quasicrystals with the one-dimensional
case \cite{Bruijn0,Bruijn6} studying so-called Beatty sequences,
\begin{equation}
\lfloor(n+1)\theta+\gamma\rfloor-\lfloor n\theta+\gamma\rfloor,\quad
\lceil(n+1)\theta+\gamma\rceil-\lceil n\theta+\gamma\rceil,
\quad (n\in \mathbb Z),
\end{equation}
with $\theta$ and $\gamma$ some constants. These sequences are in direct
correspondence with generalizations of the Fibonacci lattice and also
play a role in higher-dimensional tilings of Penrose type. They are
differences of two sequences of the type discussed in the subsection on
Kronecker's theorem.

Next, de Bruijn went to the two-dimensional Penrose case introducing the
pentagrid method \cite{Bruijn1,Bruijn2}, which he later generalized
to the multigrid method \cite{Bruijn9}. The multigrid or $n$-grid is
a collection of $n$ sets of equidistant parallel lines, with lines of
different sets making angles that are multiples of $\pi/n$. The tiling
then is a dualization of the multigrid, assigning to each intersection
of lines a rhombus with all sides perpendicular to one of the
intersecting lines. The sides of all rhombi are of equal length.

More generally, de Bruijn constructed also higher-dimensional
versions building quasicrystals of parallellotopes \cite{Bruijn9},
thus completing the proof that the dual of a multigrid is a tiling.

It should be noted that Gardner, in his 1977 Scientific American
article \cite{Gardner}, had already discussed the ``Conway worms''
discovered by Conway. In the rhombus version of the Penrose tiling,
a Conway worm is a band that passes through opposite sides of a sequence
of rhombi of the tiling. There are five sets of nonintersecting worms that
correspond one-to-one to the pentagrid of de Bruijn. 

We should also note Ammann's observation that the worms could
be replaced by a quasigrid of five sets of parallel lines that are
not equidistant, but their interdistances rather form a Fibonacci pattern.
This is now called the Ammann quasigrid made up of five grids of Ammann
bars \cite{Gardner2,GrSh}. The connection with the pentagrid of de Bruijn
was explicitly worked out and proved by de Bruijn himself in \cite{Bruijn8}.
However, the topological deformations needed for the proof are such that
one would not want to treat the two five-grid pictures as interchangeable
in actual more-involved calculations on Penrose tilings like the one
discussed in  section \ref{PIM}.

The Fourier transforms of the tilings were treated by de Bruijn in
\cite{Bruijn3,Bruijn5}, with earlier results for the regular Penrose
tiling worked out numerically by Mackay \cite{Mackay1,Mackay2}.
For the infinite tiling this is an infinite sum of Dirac delta functions,
corresponding to what physicists call the Bragg diffraction peaks.

In \cite{Bruijn4} de Bruijn made the observation that for a given infinite
Penrose tiling one can find for any number $\varepsilon>0$ two vertices
so that all other infinite Penrose tilings with the same two vertices have
all but a fraction $\varepsilon$ of their vertices in common with the
original tiling.

In \cite{Bruijn7} de Bruijn made a study of the inflation-deflation
rules of the Penrose tilings providing details left out by Gardner
\cite{Gardner}.

In several of the aforementioned papers de Bruijn also addressed the
Cut-and-Project Method. The simplest example is the one-dimensional
Fibonacci lattice seen above. It derives from the two-dimensional
square lattice by cutting a narrow band out of the lattice and
projecting all lattice points within the band. More precisely, choose
a direction with slope given by the golden ratio and choose a properly
sized window perpendicular to that direction. Then project all lattice
points within the window to the sloped line. This generalizes to $n$
and $m$ dimensions (5 and 3 for Penrose tilings) \cite{KN}.


\section{Paul Steinhardt ``to the rescue"}

While Shechtman was fighting referees and particularly Linus Pauling,
he got a big boost from the work of Steinhardt and coworkers.

Paul Steinhardt is a cosmologist, originally at the University of
Pennsylvania, but now at Princeton. He wrote many papers on such topics
as Inflationary Cosmology and Cyclic Cosmology.

Up to 1984 quasiperiodic tilings was mainly a topic for mathematicians.
Together with Levine, Socolar and others he introduced the works of
Penrose, de Bruijn, Mackay, and later Gummelt, to the physics community,
while extending these works in various directions \cite{St0,SoStL,Ste}.

Many quasicrystals have been synthesized. The obvious question was,
``Do quasicrystals occur in nature?" This has been answered in the
affirmative as a first natural quasicrystal has been discovered
in a museum in Florence, Italy. It came from the Khatyrka River in
Kamchatka, Russia \cite{Nat1,Nat2,Nat3}.


\section{Alternative approaches}

\subsection{Periodic approximants}

Tsunetsugu et al.\ designed a modification of the multigrid
approach of de Bruijn that produces a converging sequence of periodic
approximants to the Penrose tiling \cite[Appendix]{TFUT}. This could
be extremely useful for numerical calculations on the Penrose tiling,
as one can often produce very accurate numerical results for a number
of approximants that may then extrapolate very well.


\subsection{Inflation rules}

One favorite way to create large tilings is by the repeated application of
certain inflation rules \cite{BGM}: Blow up all tiles and substitute each of
them by a collection of the original tiles following very precise rules.
The Tilings Encyclopedia \cite{TilingEnc} gives more than 180 examples.


\subsection{Overlapping unit cells}

Petra Gummelt designed a special decagon decorated with red and
blue patches in a particular way \cite{Gummelt} that can partially
overlap with a copy in a few ways. Two neighboring decagons must overlap red-on-red
and blue-on-blue. There are two types of possibilities A or B.
The result is a representation of a Penrose tiling in a way that may
explain how they arise from chemical interactions. ``Gummelt's Decagon
as Overlapping (Quasi) Unit Cell" has, therefore, caused much interest.

Further aspects of the decagon approach were discussed in
\cite{SJ,JS}. This is followed by a number of studies that discuss
clusters of different types of atoms and the related chemistry filling the
decagon in with atoms, thus explaining how it comes about
\cite{SJSTAT,St,LR,JS2}. Even comparisons with experiments are made
in some of these papers \cite{SJSTAT,ASTTSJ}.
Other papers discuss the partial relaxation of the matching
rules implied by the Gummelt decagon, so that one can construct random
tilings \cite{Gummelt2,GR,RG}.

Overlapping three-dimensional versions have been constructed since,
see \cite{APoverlap,APnonreg} for a decagonal example. Using de Bruijn's
five-dimensional lattice projected down to three dimensions, rather
than two for the Penrose tiling, we constructed a quasicrystal, periodic
in one direction and quasiperiodic in the other two. It is made up of
overlapping polytopes with 22 external vertices and 4 internal ones.
But the polytopes in that model do not yet have a Gummelt-type
decoration enforcing matching rules.

The 1985 paper of Shechtman and Blech \cite{SB} used non-overlapping icosahedra.
This gave a description of the structure, but
no complete understanding how it could come about.


\section{Pentagrid Ising Model and Other Integrable Models \label{PIM}}

\subsection{Planar Ising model}

One area of physics, for which the de Bruijn pentagrid construction has been
applied, is integrable models of statistical mechanics on quasicrystals. As
an example we shall discuss those aspects of our work on the (zero-field)
pentagrid Ising model \cite{APpenta} that relate to de Bruijn's work. Ising models
on Penrose tilings belong to the class of planar Ising models. It is well-known
that such Ising models in zero magnetic field can be solved in principle using
Pfaffians or fermion methods \cite{GH,MW}, though in practice it can become
very cumbersome. We shall see that the pentagrid allows introducing an
extra integrability structure facilitating detailed calculations.

A planar Ising model is defined on a planar graph, with spin variables
$\sigma_j=\pm1$ on each vertex $j$ and interaction energy
$-J_{j,j'}\sigma_j\sigma_{j'}$ associated with each edge $\langle j,j'\rangle$.
In addition, each spin $\sigma_j$ interacts with the magnetic field, $B$ in
suitable units, so that the total interaction energy becomes
\begin{equation}
\mathcal{H}=-\sum_{\langle j,j'\rangle}J_{j,j'}\sigma_j\sigma_{j'}-B\sum_j\sigma_j,
\end{equation}
with the first sum over all edges $\langle j,j'\rangle$ of the graph and
the second sum over all $\mathcal{N}$ vertices (sites) $j$.

In equilibrium statistical mechanics one introduces the Boltzmann-Gibbs
probability distribution
\begin{equation}
\rho(\{\sigma\})=\frac{\mathrm{e}^{-\beta\mathcal{H}}}{Z},\qquad
Z=\sum_{\{\sigma\}}\mathrm{e}^{-\beta\mathcal{H}},\qquad
\beta=\frac1{k_{\mathrm{B}}T},
\label{partfun}\end{equation}
where $\{\sigma\}$ stands for all $\mathcal{N}$ values $\sigma_j=\pm1$,
($j=1\ldots\mathcal{N}$), $Z$ denotes the partition function
(Zustandssumme in German) providing the normalization of $\rho$, while
$T$ is the absolute temperature and $k_{\mathrm{B}}$ Boltzmann's constant.

Some physical quantities of particular interest are the free energy per site
\begin{equation}
f=-\frac1{\beta\mathcal{N}}\ln\sum_{\{\sigma\}}
\exp\Big(\beta\sum_{\langle j,j'\rangle}
J_{j,j'}\sigma_j\sigma_{j'}\Big)=
-\frac1{\beta\mathcal{N}}\ln Z
\label{freeen}\end{equation}
and the pair correlation of spins on sites $k$ and $l$,
\begin{equation}
\langle\sigma_k\sigma_l\rangle=\frac
{\displaystyle{\sum_{\{\sigma\}}\sigma_k\sigma_l
\exp\Big(\beta\sum_{\langle j,j'\rangle}
J_{j,j'}\sigma_j\sigma_{j'}\Big)}}
{\displaystyle{\sum_{\{\sigma\}}
\exp\Big(\beta\sum_{\langle j,j'\rangle}
J_{j,j'}\sigma_j\sigma_{j'}\Big)}}=
\sum_{\{\sigma\}}\rho(\{\sigma\})\sigma_k\sigma_l,
\label{paircorr}\end{equation}
both in zero magnetic field, $B=0$. In most physical applications
one is interested in very large systems ($\mathcal{N}\sim 10^{23}$),
so that one may just as well consider the thermodynamic limit
$\mathcal{N}\to\infty$. For the quasiperiodic case of
a ferromagnetic ($J_{j,j'}>0$) Ising model on a Penrose tiling,
the bulk properties are uniquely determined in this limit, irrespective
of what happens at the boundaries. Another physical quantity is the single
spin correlation $\langle\sigma_{\mathbf r}\rangle$ being evaluated in
infinitesimal positive magnetic field (if all $J_{j,j'}\ge0$), i.e. 
\begin{equation}
\langle\sigma_k\rangle=\lim_{B\downarrow0}\lim_{\mathcal{N}\to\infty}\frac
{\displaystyle{\sum_{\{\sigma\}}\sigma_k
\exp\Big(\beta\sum_{\langle j,j'\rangle}
J_{j,j'}\sigma_j\sigma_{j'}+\beta B\sum_j\sigma_j\Big)}}
{\displaystyle{\sum_{\{\sigma\}}
\exp\Big(\beta\sum_{\langle j,j'\rangle}
J_{j,j'}\sigma_j\sigma_{j'}+\beta B\sum_j\sigma_j\Big)}}.
\label{magn}\end{equation}
This is positive (nonzero) for temperatures below the critical
temperature, $T<T_{\mathrm{c}}$. The order of limits is important \cite{Yang},
as in the other order we have $B=0$ for finite $\mathcal{N}$, in which case the
symmetry $\sigma_j\to-\sigma_j$ for all $j$ negates $\langle\sigma_k\rangle$ 
implying $\langle\sigma_k\rangle\equiv0$.


\subsection{Star-triangle (Yang--Baxter) equation}

\begin{figure}[htbp]
\begin{center}
\includegraphics[height=7.4cm]{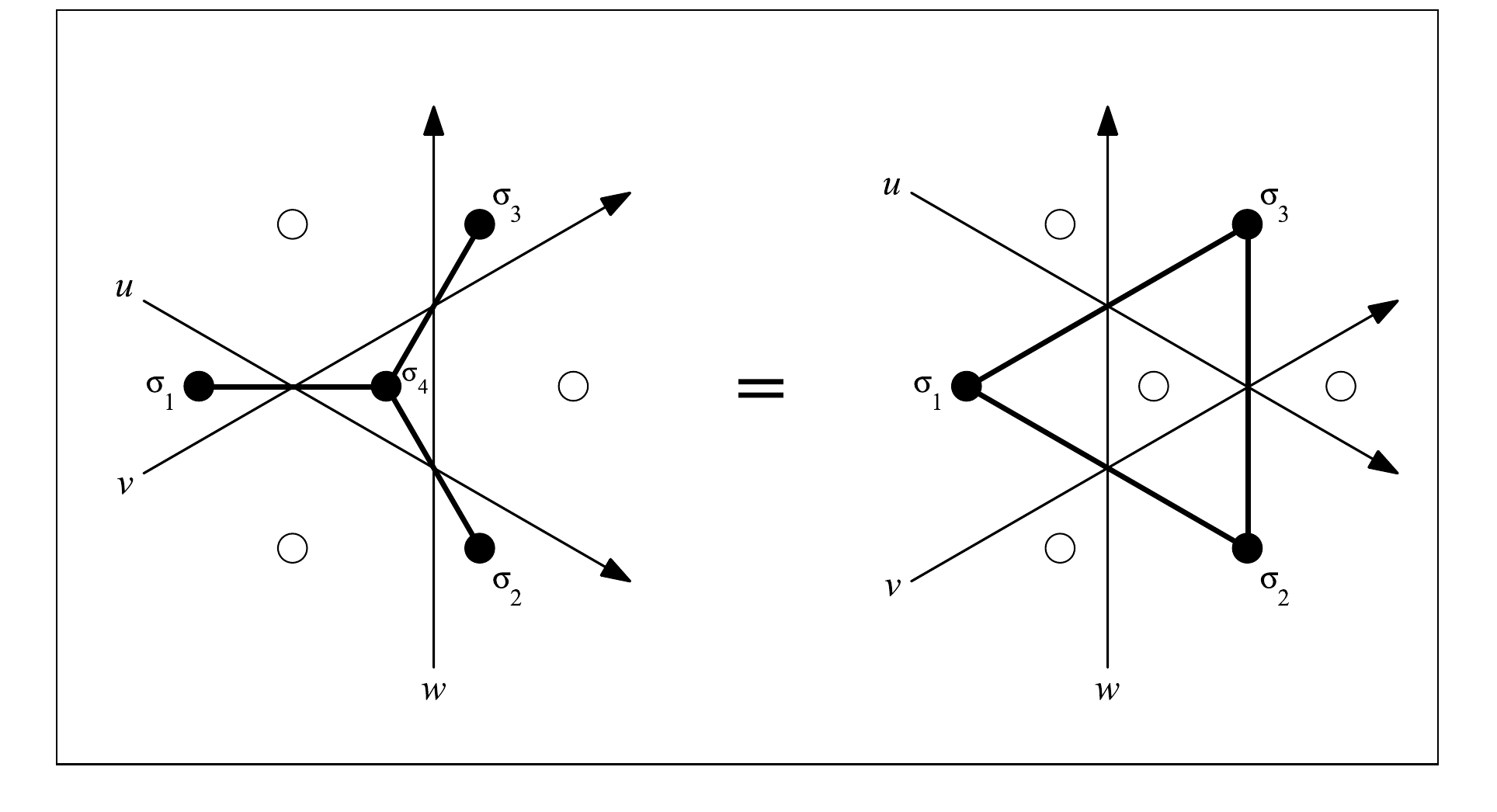}
\includegraphics[height=5cm]{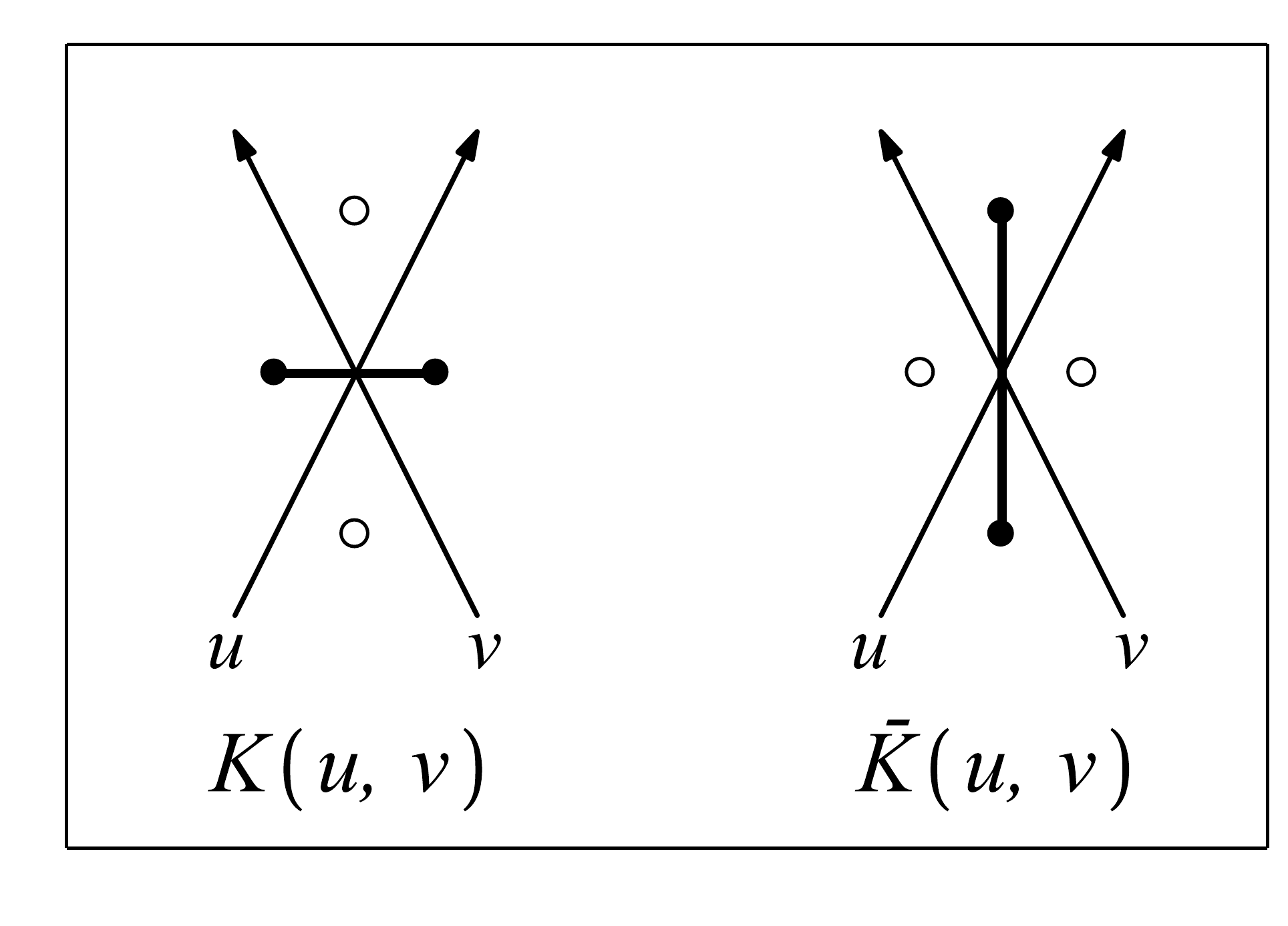}
\end{center}
\caption{The star-triangle (Yang--Baxter) equation for the Ising
model represented as a picture. Solid circles represent Ising spins
coupled by fat lines. Open circles represent spin sites of the dual
Ising model. The thin oriented lines carry the rapidity variables.
The bottom picture shows the two assignments of $\beta J$ by either
function $K(u,v)$ or $\bar K(u,v)$.}
\label{fig2}\end{figure}

For a subclass of Ising models more can be solved
using an underlying Yang--Baxter equation \cite{YB}, which becomes
a star-triangle relation in this case, namely,
\begin{eqnarray}
&\displaystyle\sum_{\sigma_4=\pm1}\exp\Big(\bar K(u,v)\sigma_1\sigma_4+
K(u,w)\sigma_2\sigma_4+\bar K(v,w)\sigma_3\sigma_4\Big)\nonumber\\
&\displaystyle\quad=R(u,v,w)\exp\Big(K(v,w)\sigma_1\sigma_2+
\bar K(u,w)\sigma_1\sigma_3+K(u,v)\sigma_2\sigma_3\Big),
\label{ST}
\end{eqnarray}
with $K_{j,j'}\equiv\beta J_{j,j'}=K(u,v)$ or $\bar K(u,v)$,
depending on the orientation of the edge $\langle j,j'\rangle$
with respect to two oriented lines crossing the edge, while these lines
carry ``rapidity'' variables $u$ and $v$, see also figure \ref{fig2}.
Also, $R(u,v,w)$ is a factor independent of the values of $\sigma_1$,
$\sigma_2$ and $\sigma_3$.

Equation (\ref{ST}), illustrated by figure \ref{fig2}, expresses the fact that
for an Ising model so parametrized the partition function $Z$ does not change,
if any given rapidity line passes through the intersection of two other rapidity
lines. Therefore, such models are called $Z$-invariant by Baxter \cite{BaxZI}.
Many other solutions of Yang--Baxter equations exist that are not Ising-like.
For example, Baxter treats the $Z$-invariant eight-vertex model in the same
paper \cite{BaxZI}, a model for which the Yang--Baxter equation is still
pictured by figure~\ref{fig2}, provided one couples each pair of spins connected
by a fat line with the two dual spins on both sides of that line. See, e.g.,
\cite{YB} for various appearances of the Yang--Baxter equation.

\begin{figure}[htbp]
\begin{center}
\includegraphics[height=7cm]{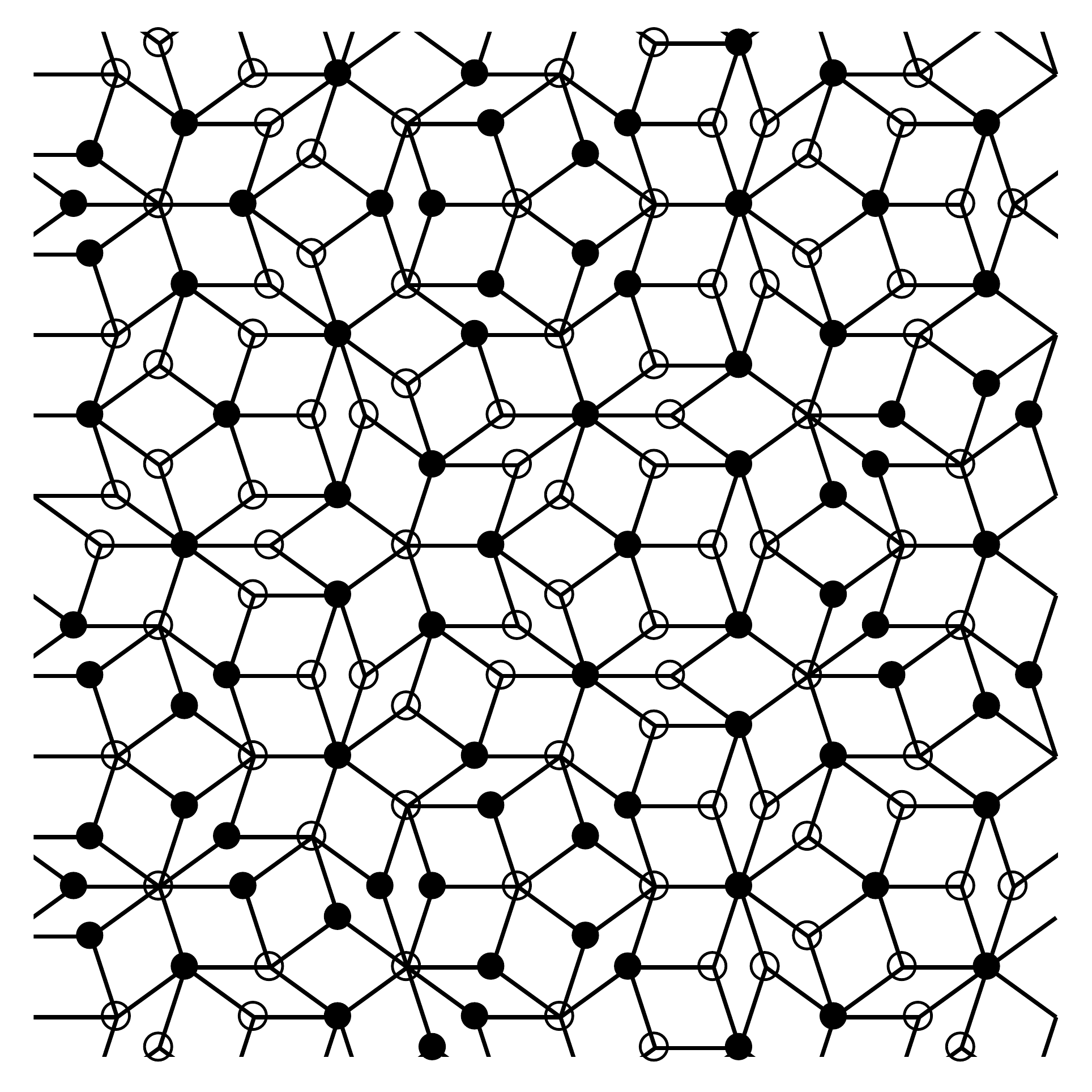}
\end{center}
\caption{Part of a regular Penrose tiling of fat and skinny rhombi.
The graph is bipartite, so that the vertices can be colored
alternatingly black and white, dividing the tiling into two (quasi-)sublattices.}
\label{fig3}\end{figure}

Korepin \cite{K1,K2,AK1,AK2} pioneered the eight-vertex model on a
Penrose tiling. Now the rapidity variables, the free parameters in the
Yang--Baxter equation, live on the grid lines of de Bruijn's pentagrid.
The model has four-spin interactions for each quadruple of vertices
belonging to a rhombus of the tiling, see figure \ref{fig3}.
This allowed Korepin to calculate the free energy per site of the model
on an infinite Penrose tiling following Baxter \cite{BaxZI,BaxBook}
and also to apply Baxter's ``unwieldy'' multiple integral formula
\cite{BaxZI} for the pair correlation in the special case that there is
only two-spin interactions across all diagonals of the rhombi. This special
case factorizes into two independent pentagrid Ising models.


\subsection{The pentagrid Ising model}

\begin{figure}[htbp]
\begin{center}
\includegraphics[height=6cm]{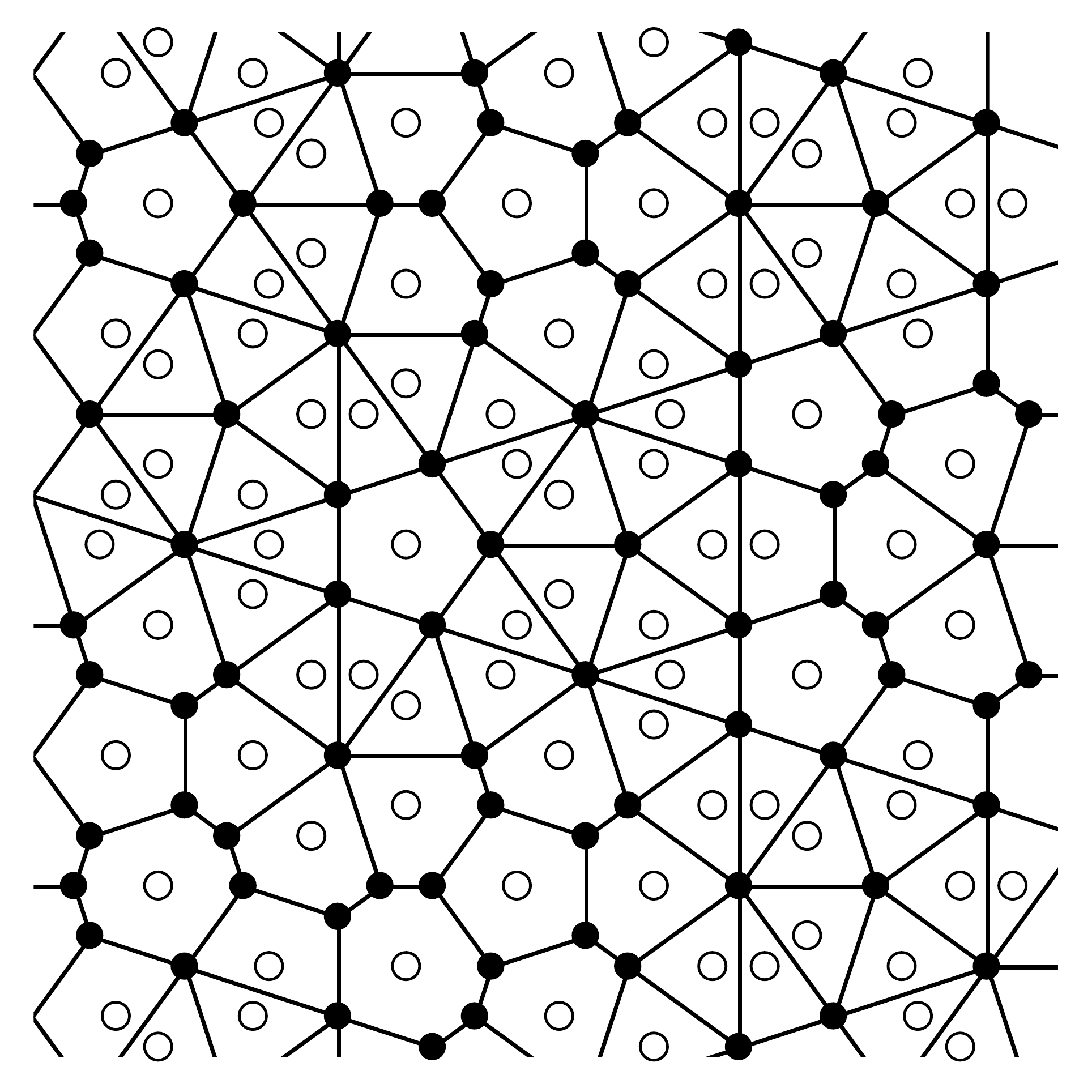}
\includegraphics[height=6cm]{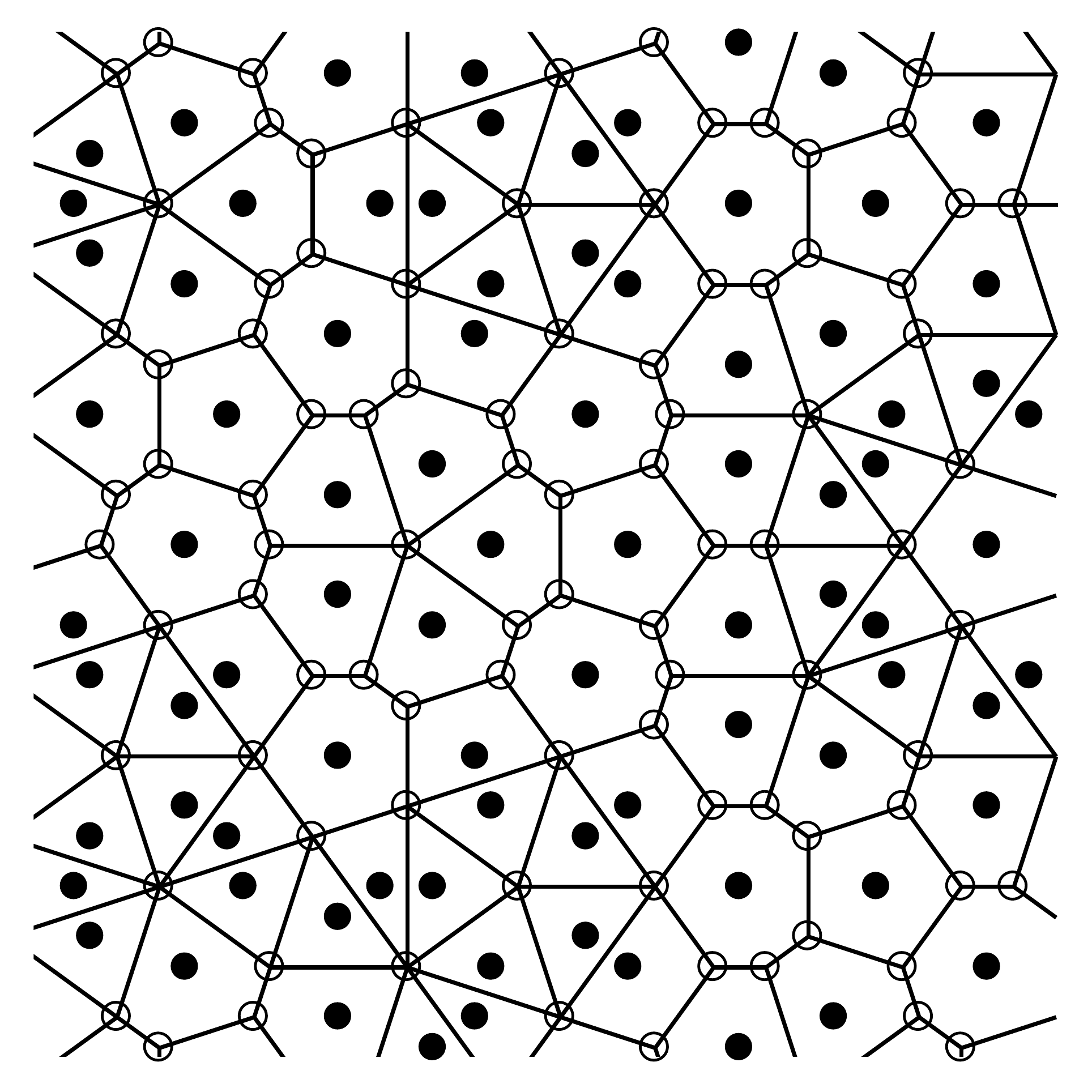}
\end{center}
\caption{On the left: The pentagrid Ising model with spins on the
black lattice, interacting along diagonals of the rhombi. On the
right: The dual pentagrid Ising model built similarly with
spins on the white lattice.}
\label{fig4}\end{figure}

If we divide the vertices of a Penrose rhombus tiling into two sublattices
as shown in figure \ref{fig3}, we can put an Ising model on one sublattice
with Ising interactions along the diagonals of the rhombi. The other
sublattice is the dual lattice of the first one, on which one can put an
Ising model at the dual temperature. We have called this model, shown in
figure \ref{fig4}, the pentagrid Ising model \cite{APpenta}. This last paper
consists of three parts, namely,
\begin{itemize}
\item an algebraic recursion scheme to systematically calculate results of
pair correlations as functions of the rapidities on pentagrid lines
``passing between'' the pairs of spins, without doing Baxter's integrals
over an ever-increasing number of variables,
\item a new method to determine the joint probability of two distant
local environments in the pentagrid, to connect those pair
correlations with actual positions of spins,
\item the explicit numerical calculation of the Fourier transform
($q$-dependent susceptibility or structure function) of the
pentagrid-Ising-model pair correlation.
\end{itemize}
We shall next briefly describe these three items, as they relate to
de Bruijn's pentagrid construction, without going into the full details
of \cite{APpenta}.

\begin{figure}[htbp]
\begin{center}
\includegraphics[height=6cm]{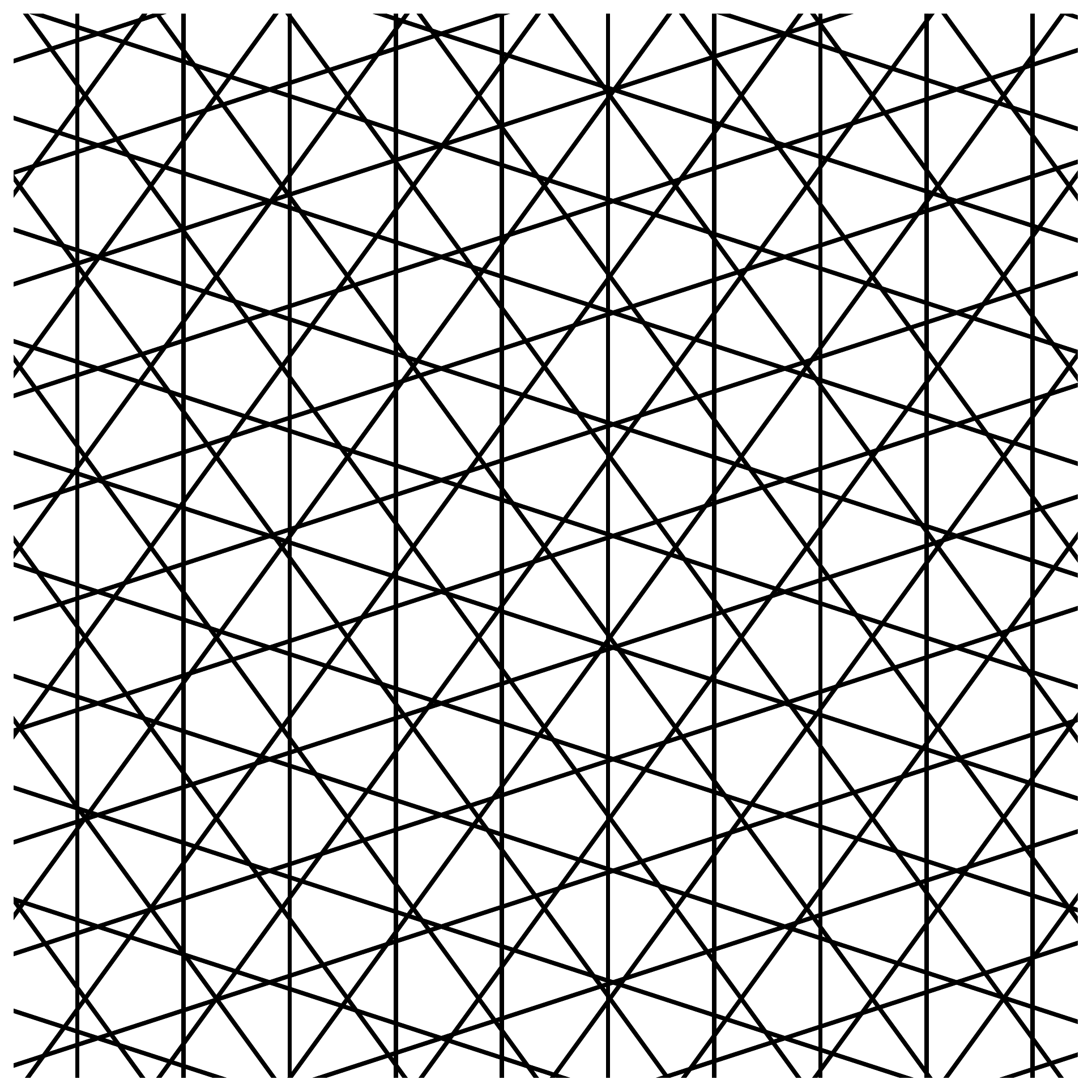}
\includegraphics[height=6cm]{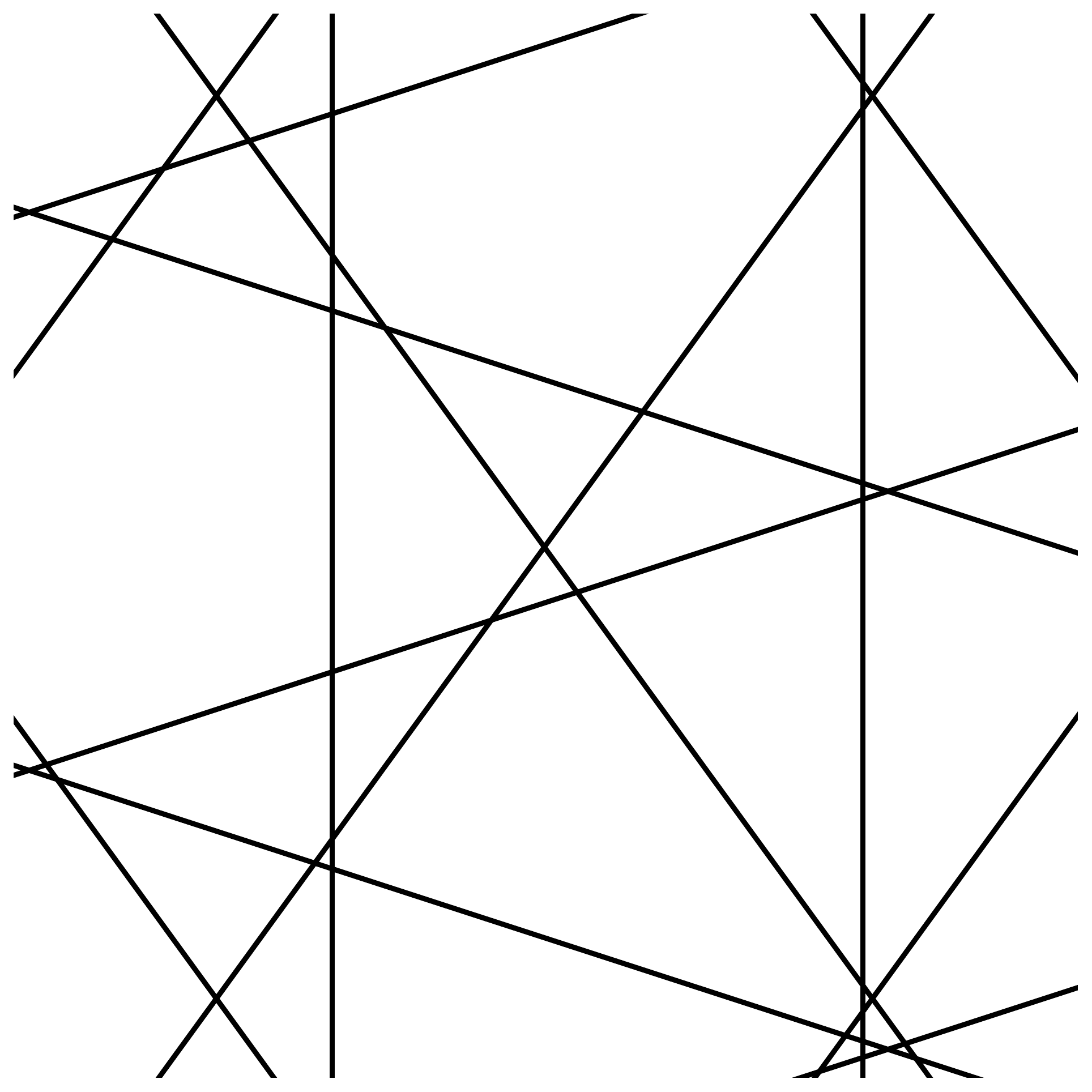}
\end{center}
\caption{On the left: Part of the pentagrid. No three lines pass
through a common point, even though pairs of intersections may approach
arbitrarily close. On the right: The central ``parallelogram'' from
lines of two grids with one of 24 topologically different
configurations of lines from the other three grids inside. Note that
multiple intersections indeed do not take place.}
\label{fig5}\end{figure}

In order to define the integrable pentagrid Ising model we start with
a regular pentagrid \`a la de Bruijn \cite{Bruijn1,Bruijn2}---five grids
of equidistant parallel lines making angles of multiples of $\pi/5$ with
one another and shifted such that no three lines meet in a common
intersection, see figure \ref{fig5}. These lines, carrying the
rapidity variables, correspond to curved lines in the Penrose tiling, with
each line passing through opposite sides of a sequence of rhombi in
figure \ref{fig3}, thus identifying a grid line of the pentagrid with
a ``Conway worm'' \cite{Gardner}---band of rhombi with each rhombus
connected to two other ones at opposite edges.

\begin{figure}[htbp]
\begin{center}
\includegraphics[height=5cm]{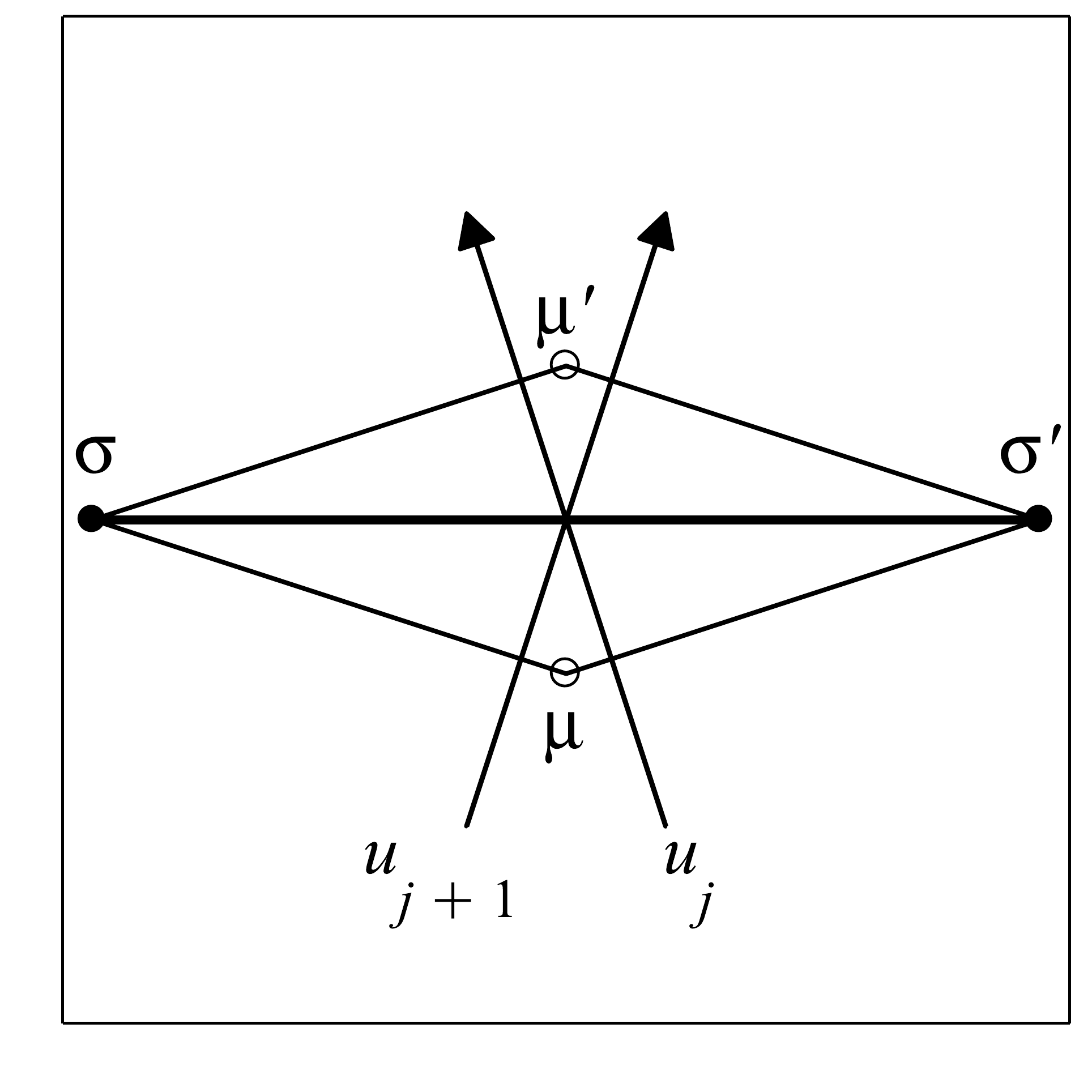}
\includegraphics[height=5cm]{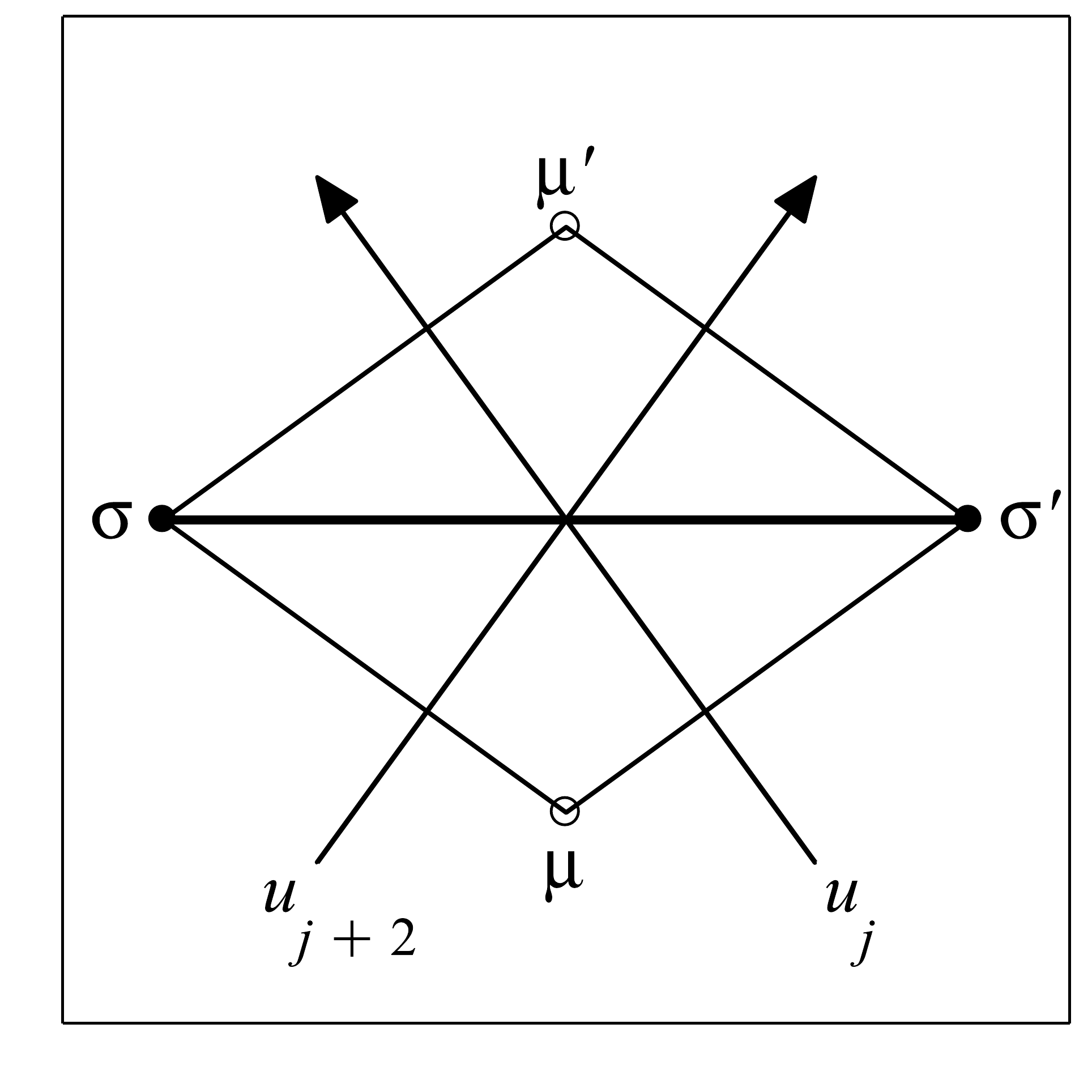}
\includegraphics[height=5cm]{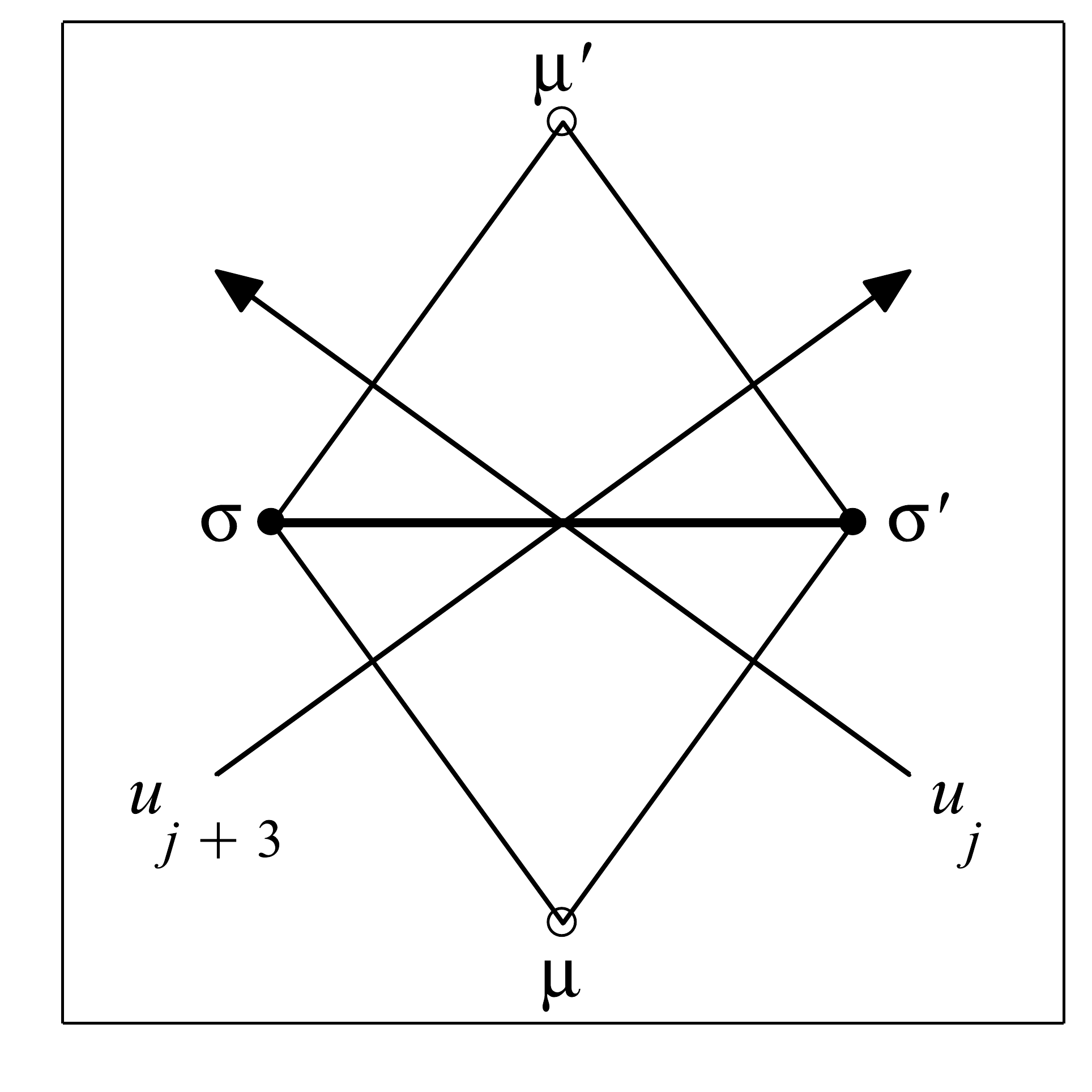}
\includegraphics[height=5cm]{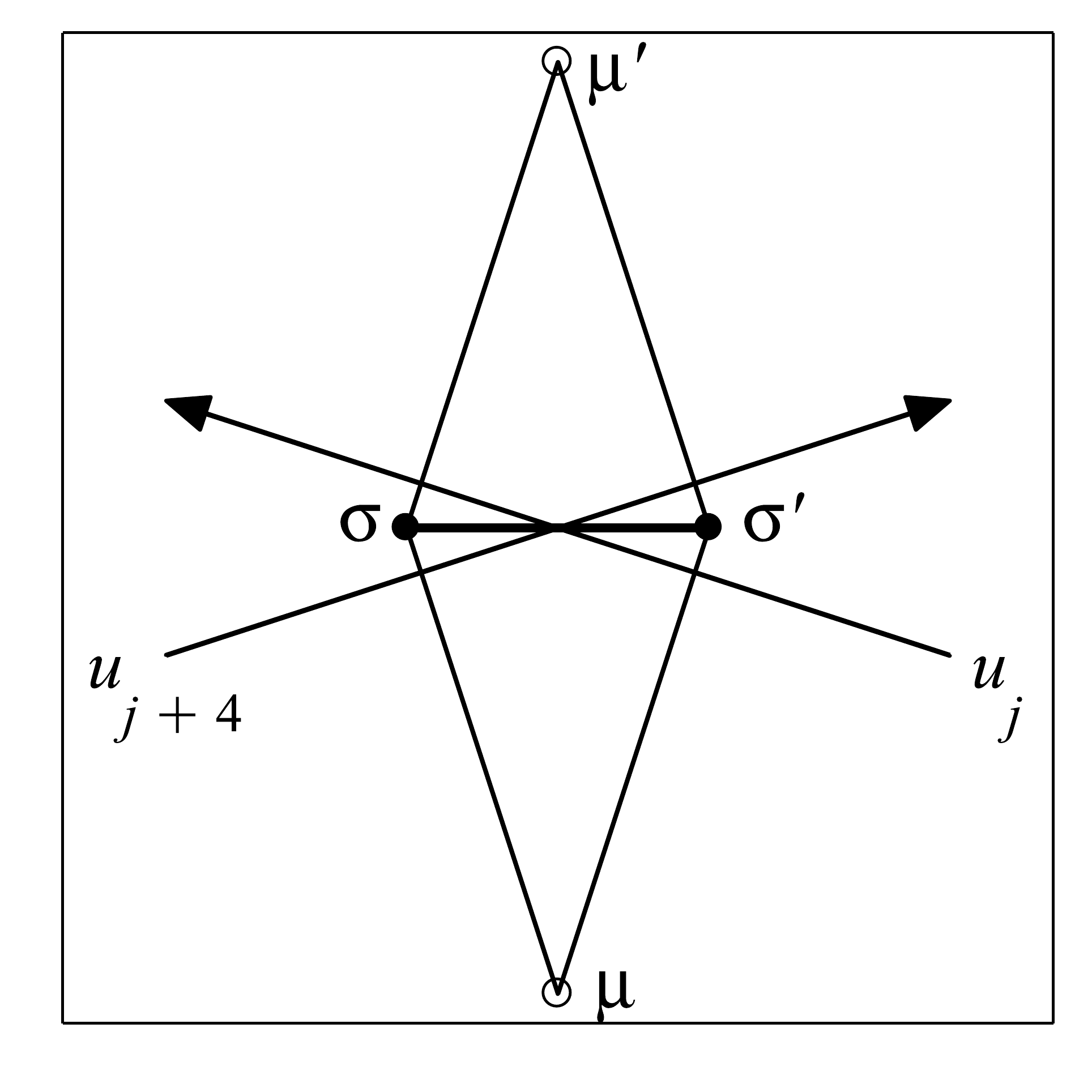}
\end{center}
\caption{The four different ways to assign a rhombus to a given
intersection of the pentagrid, with four different-size diagonals
connecting the black vertices.}
\label{fig6}\end{figure}

Grid lines can only intersect with angles of $\pi/5$ or $2\pi/5$. Following
de Bruijn \cite{Bruijn1,Bruijn2}, to each such intersection we assign a skinny
or fat rhombus, respectively with their sides perpendicular to the grid lines,
as is shown in figure \ref{fig6}.

\subsection{Choice of Ising interactions}

We want to choose the parameters of the interactions such that the
star-triangle equations hold, that the model has a five-fold symmetry axis,
and that the interactions are weaker if the separation of spins is larger.

The star-triangle equation imposes a parametrization in terms of elliptic
functions of modulus $0\leqslant k\leqslant1$, $k'=\sqrt{1-k^2}$. This $k$
relates to the temperature, with $k=1$ being the critical case.
For the two choices for $\beta J=K(u,v)$ or $\bar K(u,v)$ in figure \ref{fig2}
we must choose, in the low-temperature case with $k\leqslant1$ a
temperature-like variable,
\begin{eqnarray}
&&\sinh\big(2K(u,v)\big)=\mathrm{sc}(u-v,k')
=k^{-1}\mathrm{cs}\big(\mathrm{K}(k')-u+v,k'\big),
\label{Ks}\\
&&\sinh\big(2{\bar K}(u,v)\big)=k^{-1}\mathrm{cs}(u-v,k')
=\mathrm{sc}\big(\mathrm{K}(k')-u+v,k'\big),
\label{Ksb}
\end{eqnarray}
with ${\mathrm K}(k')$ the complete elliptic integral of the first kind of
modulus $k'$ and $\mathrm{sc}=\mathrm{sn}/\mathrm{cn}$ and
$\mathrm{cs}=1/\mathrm{sc}$ Jacobi elliptic functions. In the
high-temperature case, with $k^{-1}\geqslant1$ temperature-like, we choose
\begin{eqnarray}
&&\sinh\big(2K(u,v)\big)=k\,{\rm sc}(u-v,k')
={\rm cs}\big({\rm K}(k')-u+v,k'\big),
\label{Ksd}\\
&&\sinh\big(2{\bar K}(u,v)\big)={\rm cs}(u-v,k')
=k\,{\rm sc}\big({\rm K}(k')-u+v,k'\big).
\label{Ksdb}
\end{eqnarray}

Next we must choose the rapidities, such that the model's interaction
strengths are rotational invariant, only depending on the separations of
spins. This is easily achieved noting that (\ref{Ks}) through (\ref{Ksdb})
only depend on $k$ and the difference of rapidities $u-v$. Each of the five
grids of the pentagrid can be oriented in two directions. Let for the $j$th
oriented grid, $\mathrm{grid}_j$, the arrow of each grid line point in
the direction defined by angle
\begin{equation}
\phi_j=\angle\,\mathrm{grid}_j=\phi_0-j\pi/5,\quad\mbox{and let}\quad
u_j=j{\rm K}(k')/5,
\label{grangle}\end{equation}
with $j$ defined modulo 10. It is clear that then $\mathrm{grid}_{j\pm5}$
is the same grid as $\mathrm{grid}_j$, but with the arrows reversed,
and $u_{j\pm5}=u_j\pm{\rm K}(k')$. This condition for the change of rapidity
with arrow reversion originates from \cite[page 50]{APZI}.

The four pictures in figure \ref{fig6} represent $\beta J=K(u_{j+l},u_j)$,
for $l=1,\ldots,4$, see also figure \ref{fig2}. Inverting the arrow of
the $u_j$-rapidity, replacing $u_j$ by $u_{j+5}$, we obtain
$\beta J=\bar K(u_{j+5},u_{j+l})$, which is the same, as can be seen
comparing (\ref{Ks}) and (\ref{Ksb}), or (\ref{Ksd}) and (\ref{Ksdb}),
for the below or above critical temperature cases. We conclude
that $\sinh(2\beta J)=\mathrm{sc}\big(l\,\mathrm{K}(k'),k'\big)$, or
$k\,\mathrm{sc}\big(l\,\mathrm{K}(k'),k'\big)$ for $l=1,\ldots,4$,
the extra factor $k$ being needed above the critical temperature. Indeed,
the longer the diagonal the weaker the interaction. We note that Choy \cite{Choy}
made essentially the same choice of Ising interactions, specializing the
Penrose eight-vertex model and the free energy results of Korepin \cite{K1,K2},
but he did not derive new results for correlation functions.


\subsection{Pair correlations in terms of rapidities}

The traditional way to calculate pair correlation function (\ref{paircorr})
in infinite planar Ising models is using either fermion (Clifford algebra) methods
or equivalent dimer problems \cite{GH,MW}. If the spin pair lies on an axis
of reflection symmetry, this leads to Toeplitz determinants, but in general
the computation can become quite cumbersome.

In many cases, the work can be greatly reduced with the help of quadratic
identities for the pair correlations \cite{jhhp80}, enabling one to exactly
calculate all pair correlation functions one-by-one recursively. These identities
can be seen as a general Wick theorem for fermions or as a compound Pfaffian
theorem \cite{PCQN}. Further simplification occurs if the model has an
underlying $Z$-invariance \cite{BaxZI,APZI}. Since the pentagrid Ising model
is defined through rapidities on the pentagrid, this applies \cite{APpenta}
also for this case.

Let us explain this in more detail. Consider a pair of spins in the bulk
of a very large but finite $Z$-invariant Ising model. All edges of the model
correspond to intersections of pairs of rapidity lines. Bend, if necessary,
the rapidity lines such that there are no further intersections as they go
off to infinity. Then the edges of the Ising model are in one-to-one
correspondence with the intersections of the rapidity lines. We can now
play Baxter's game of moving rapidity lines to or from infinity, making
changes on the boundary of the Ising model that cause perimeter-to-area
effects in the free energy per site that vanish in the infinite system limit.%
\footnote{The pair correlation has high- and low-temperature expansions
that are uniformly convergent within radii of convergence (expected to
correspond to modulus $k=1$), with more and more coefficients taking their
limiting values as the boundary moves away from the two spins with increasing
system size.}

We can bend a rapidity line that goes through the bulk of the model, but
does not pass between the pair of spins considered, such that it now
intersects outside the model boundary with a sequence
of other rapidity lines. Then, applying the star-triangle equation of
figure \ref{fig2} several times, we can move that rapidity line to the
boundary. Similarly, we can introduce a new rapidity line at the boundary
and move it to go near the spin pair considered. However, we cannot move
rapidity lines that cross between spins in and out from there; we can
only change their ordering.

Consequently, the pair correlation cannot depend on rapidity variables
of rapidity lines that do not cross between the spin pair. It does
depend on the elliptic modulus and the set of rapidity variables
on the even number, say $2n$, of lines that do cross between the spins.
With Baxter \cite{BaxZI} we can now introduce universal functions
$g_{2n}(\{u_1,\ldots,u_{2n}\},k)$, for $n=0,1,2,\ldots$, and
similarly $g^{\ast}_{2n}$ for the dual Ising model (with spins on
the open circles rather than the solid circles in figures \ref{fig3}
and \ref{fig4}). More explicitly,
\begin{equation}
\langle\sigma_i\sigma_j\rangle=g_{2n}(\{u_1,\ldots,u_{2n}\},k),
\quad\mbox{or}\quad g(u_1,\ldots,u_{2n})\quad\mbox{for short},
\label{guniv}\end{equation}
where $u_1,\ldots,u_{2n}$ is any ordering of the rapidities of the
rapidity lines passing between sites $i$ and $j$ in the same direction.
As said below (\ref{grangle}), changing the direction of rapidity line $k$
results in the replacement $u_k\to u_k\pm{\rm K}(k')$.
Obviously, $g_0\equiv g^{\ast}_0\equiv1$, as $\sigma_i^{\,2}=(\pm1)^2=1$.

With the notation of (\ref{guniv}) the quadratic identities of \cite{jhhp80}
become
\begin{eqnarray}
&&{\rm sc}(u_2\!-\!u_1,k'){\rm sc}(u_4\!-\!u_3,k')\nonumber \\
&&\quad\times\big\{g(u_1,u_2,u_3,u_4,\cdots)g(\cdots)\!
-\!g(u_1,u_2,\cdots)g(u_3,u_4,\cdots)\big\}\nonumber \\
&&+\big\{g^\ast(u_1,u_3,\cdots)g^\ast(u_2,u_4,\cdots)\!
-\!g^\ast(u_1,u_4,\cdots)g^\ast(u_2,u_3,\cdots)\big\}=0,\nonumber\\
\cr
&&k^2{\rm sc}(u_2\!-\!u_1,k'){\rm sc}(u_4\!-\!u_3,k')\nonumber \\
&&\quad\times\big\{g^\ast(u_1,u_2,u_3,u_4,\cdots)g^\ast(\cdots)\!
-\!g^\ast(u_1,u_2,\cdots)g^\ast(u_3,u_4,\cdots)\big\}\nonumber \\
&&+\big\{g(u_1,u_3,\cdots)g(u_2,u_4,\cdots)\!
-\!g(u_1,u_4,\cdots)g(u_2,u_3,\cdots)\big\}=0,\label{Toda}
\end{eqnarray}
where all ``$\cdots$'' stand for the same (but arbitrary) even set
of $u_j$'s, see \cite{APO1,APO2}. From these one can systematically derive
all $g_{2n}$ and $g^{\ast}_{2n}$ iteratively with increasing $n$, provided
one knows these for the two cases with all or all but one of the $u_j$'s
equal. For the pentagrid Ising model paper \cite{APpenta}, these special
cases were evaluated by iteration from determinant formulae in \cite{APO1}.
Now there is a more efficient way, discovered by Witte \cite{Witte} and
described in detail in \cite[section 3]{CGMP}.

Finally, the single-spin correlation (\ref{magn}) cannot depend on any
rapidity variable. Therefore \cite{BaxZI,BaxBook},
\begin{equation}
\langle\sigma_i\rangle\equiv{k'}^{1/4}\quad\mbox{for all sites }i,
\label{orderpar}\end{equation}
below criticality, and zero above. Because of this property, it is also called
order parameter or spontaneous magnetization per site.


\subsection{Integer quintuples, indices, and probabilities\label{prob}}

So far we have not used the more technical part of de Bruijn's work on
Penrose tilings. If one only wants to know the free energy per site of
the pentagrid Ising model in the thermodynamic limit, one just needs the
idea of the pentagrid construction and the ratio of the numbers of fat and
skinny rhombi in an infinite Penrose tiling. Everything else needed is
supplied by Baxter \cite{BaxZI}. The order parameter (\ref{orderpar}) is
even a freebie in Baxter's approach.

However, once we want to know more about the magnetic structure of the
model as is contained in the pair correlation, we need to know the precise
position of spins in relation to the pentagrid. We also need a prescription
to determine if a spin is in the black (even) or in the white (odd)
sublattice of figure \ref{fig3}. Finally, we need the joint probability
of the occurrence of two local configurations that are an arbitrary
distance apart. For all these we need many technical details of
\cite{Bruijn1,Bruijn2,Bruijn0}, as is explained more fully in \cite{APpenta}.
Here we shall outline the ideas, especially the use of sections 4 and 5
of \cite{Bruijn1}.

First of all, de Bruijn introduces the five grids $G_j$, with the lines of
the $j$th grid labeled by integers $k_j$, i.e.,
\begin{equation}
G_j=\{z\in \mathbb{C}| {\rm Re} (z\zeta^{-j})
+\gamma_j=k_j, k_j\in
\mathbb{Z}\},\qquad j=0,\cdots,4.
\label{grid}
\end{equation}
where
\begin{equation}
\zeta={\rm e}^{2{\rm i}\pi/5},\qquad
\zeta+\zeta^{-1}=2\cos(2\pi/5)=\tau^{-1}={{\textstyle \frac 1 2}}(\sqrt 5-1),
\label{goldratio}
\end{equation}
in which $\tau$ is the golden ratio and the $\gamma_j$ are five real numbers
satisfying $\sum_{j=0}^4\gamma_j=0$. Grid $G_j$ corresponds to
$\mbox{grid}_{2j}$ with $\phi_0=0$ in (\ref{grangle}). One can easily
choose the $\gamma_j$ such that the pentagrid is regular, nowhere three
or more grid lines having a common intersection.

The vertices of the Penrose tiling correspond to the meshes, or faces,
of the pentagrid. These meshes are determined by a quintuple of integers,
the integer vector ${\vec K}(z)=\big(K_0(z),\cdots,K_4(z)\big)$---defined
the same for each point $z\in\mathbb{C}$ of the mesh---using
\begin{equation}
K_j(z)=\lceil{\rm Re} (z\zeta^{-j})
+\gamma_j\rceil,
\label{mesh}\end{equation}
with $\lceil x\rceil$ denoting the ceil or roof of $x$, i.e.\ the smallest
integer $\geqslant x$. It is easily seen from (\ref{grid}) and (\ref{mesh})
that whenever $z$ moves across a line of the $j$th grid, $K_j(z)$ changes
by 1 and that the meshes are in one-to-one correspondence with these
integer vectors.

A particular useful property discovered by de Bruijn \cite{Bruijn1} and
exploited in \cite{APpenta} is that the sum of the five integers satisfies
\begin{equation}
\mbox{Index}(z)\equiv\sum_{j=0}^4 K_j(z)=\mbox{1, 2, 3, or 4},
\end{equation}
by which we can distinguish if the corresponding vertex of the Penrose tiling,
\begin{equation}
f(z)=\sum_{j=0}^4 K_j(z)\zeta^j,
\end{equation}
belongs to the even ($\mbox{Index}=2,4$) or the odd ($\mbox{Index}=1,3$)
sublattice of figure \ref{fig3}, i.e., if it belongs to the Ising or the
dual Ising model in figure \ref{fig4}.

One way to systematically account for all vertices of the infinite Penrose
tiling is to divide the pentagrid space into parallelograms by
grids $G_0$ and $G_1$. Let $P(k_0,k_1)$ be the parallelogram bounded by
lines $k_0-1$ and $k_0$ of $G_0$ and $k_1-1$ and $k_1$ of $G_1$,
defined in (\ref{grid}). The lines of the other three grids can pass through
$P(k_0,k_1)$ in 24 topologically distinct configurations, with one
example shown in figure \ref{fig5} and all 24 cases in \cite[fig.~6]{APpenta}.
It is easily verified that this way $P(k_0,k_1)$ is split into 6 to 12
meshes (or faces) depending on the configuration.

It is convenient to single out one mesh of $P(k_0,k_1)$ that corresponds
to a vertex of the even sublattice of the Penrose tiling. For this purpose
we introduced the reference integer vector $(k_0,\cdots,k_4)$, which is
related to the extreme corner $z_{\mathrm{c}}$ of $P(k_0,k_1)$ where the
lines labeled by $k_0$ and $k_1$ cross. Thus we chose
\begin{eqnarray}
&&k_0=K_0(z_{\mathrm{c}}),\quad
k_1=K_1(z_{\mathrm{c}}),\nonumber\\
&&k_2=\lceil\alpha\rceil-k_0=K_2(z_{\mathrm{c}}),\quad
k_4=\lceil\beta\rceil-k_1=K_3(z_{\mathrm{c}}),
\label{k0124}\end{eqnarray}
with $\lceil x\rceil$ the roof or ceiling of $x$ and
\begin{equation}
\alpha\equiv\tau^{-1}(k_1-\gamma_1)+\gamma_0+\gamma_2,\quad
\beta\equiv\tau^{-1}(k_0-\gamma_0)+\gamma_1+\gamma_4,
\label{alphabeta}\end{equation}
which can be easily derived. However, for the last component of the
reference integer vector we chose
\begin{equation}
k_3=2-\lceil\alpha\rceil-\lceil\beta\rceil
=-\lfloor\alpha\rfloor-\lfloor\beta\rfloor
\not\equiv K_3(z_{\mathrm{c}})=\lceil-\alpha-\beta\rceil,
\label{k3}\end{equation}
with $\lfloor x\rfloor$ denoting the ``floor of $x$'', i.e.\ the largest
integer $\leqslant x$. The index of the reference integer vector is
$\sum_j k_j=2$, so that $(k_0,\cdots,k_4)$ indeed labels a mesh corresponding
to the even sublattice. On the other hand,
\begin{equation}
K_3(z_{\mathrm{c}})=\cases{\begin{array}{llr}
k_3-1 &\hbox{for}&
\{\alpha\}+\{\beta\}\geqslant 1,\\
k_3 &\hbox{for}&
\{\alpha\}+\{\beta\}<1,\end{array}}
\label{k3a}\end{equation}
using the notation $\{x\}\equiv x-\lfloor x\rfloor$ for the fractional
part of $x$.

This shows that only for $\{\alpha\}+\{\beta\}<1$ the mesh adjacent
to the extreme corner $z_{\mathrm{c}}$ belongs to the even sublattice,
its integer vector $\vec K(z_{\mathrm{c}})$ being the reference integer
vector of $P(k_0,k_1)$. This is not so if $\{\alpha\}+\{\beta\}\geqslant 1$,
in which case $\vec K(z_{\mathrm{c}})$ corresponds to an odd vertex. That
in one of the 24 configurations the reference integer vector lies
outside $P(k_0,k_1)$ is not a problem, as now
$\{\alpha\}+\{\beta\}\gtrless1$ determines the nature (odd/even) of
the mess adjacent to $z_{\mathrm{c}}$ and the corresponding vertex of
the Penrose tiling. 

The introduction of the variables $\alpha$ and $\beta$ in (\ref{alphabeta})
is one of the main tricks in \cite{APpenta}. The values of $\{\alpha\}$
and $\{\beta\}$ determine in which of the 24 configurations $P(k_0,k_1)$
is. It is straightforward to denote in the unit square of plotting
$\{\alpha\}$ versus $\{\beta\}$ the area corresponding to each
configuration. It is also clear from (\ref{alphabeta}) and Kronecker's
theorem of subsection \ref{kron}, that the points $(\{\alpha\},\{\beta\})$
are uniformly and densely distributed as $k_0$ and $k_1$ run through all
integers. The probability of a configuration becomes simply the magnitude
of an area in the unit square!

For the determination to a pair correlation with spins having positions
corresponding to meshes in $P(k_0,k_1)$ and $P(k_0+\Delta k_0,k_1+\Delta k_1)$,
we can determine the shift $(\Delta\alpha,\Delta\beta)$ from (\ref{alphabeta}).
It is easy to find the joint probability of having given configurations of the two parallelograms, as $k_0$ and $k_1$ run through all integers, but with
$\Delta k_0$ and $\Delta k_1$ kept fixed. It is now  expressed through
the areas of overlap of areas in the above unit square and another
square where the areas are shifted cyclically by $(\Delta\alpha,\Delta\beta)$.
This gives $24\times24$ probabilities.

However, for the regular Penrose tiling we do not have to consider all
$576=24^2$ possibilities separately, as we can combine many cases together.
The way \cite{APpenta} is set up, for the odd quasi-lattice we end up
with $8^2$ possibilities and for the even quasi-lattice $16^2$. We could even
prove that the $\chi({\mathbf q})$, defined in the next subsection, is
the same for both quasi-lattices. This reduction does not occur for
nonregular Penrose tilings \cite{APnonreg}, even though we may expect
these to lead to the same values of $\chi({\mathbf q})$.


\subsection{Wavevector-dependent susceptibility}

A quantity of particular interest in physics is the wavevector-dependent
susceptibility $\chi(q_x,q_y)$, which is the Fourier transform of
the connected pair correlation. More precisely,
\begin{equation}
\chi({\mathbf q})=\beta
\lim_{{\mathcal L}\to\infty}{\frac{1}{{\mathcal L}}}
\sum_{\mathbf r}\sum_{\mathbf r'} {\mathrm e}^{{\mathrm i}{\mathbf q}
\cdot({\mathbf r'}-{\mathbf r})}
\big[{\langle\sigma_{\mathbf r}\sigma_{\mathbf r'}\rangle}-
\langle\sigma_{\mathbf r}\rangle\langle\sigma_{\mathbf r'}\rangle\big],
\label{chi}\end{equation}
where $\mathbf{q}=(q_x,q_y)$ while $\mathbf{r}$ and $\mathbf{r}'$
are the physical positions of the vertices on which the spin variables
live. The sums are done over all spin positions in finite patches
$\mathcal L$ of the Penrose tiling and the limit is taken in which the
patch becomes the entire infinite tiling. ``Connected'' means that we
subtract the contribution of spontaneous ordering, causing Bragg-like
Dirac delta-function peaks to occur below the critical temperature.

The resulting $\chi({\mathbf q})$ is a continuous function, also
called the structure function giving information about the magnetic
structure. It is not to be confused with
\begin{equation}
\mathcal{F}({\mathbf q})=
\lim_{{\mathcal L}\to\infty}{\frac{1}{{\mathcal L}}}
\sum_{\mathbf r}\sum_{\mathbf r'} {\mathrm e}^{{\mathrm i}{\mathbf q}
\cdot({\mathbf r'}-{\mathbf r})}
\langle\sigma_{\mathbf r}\rangle\langle\sigma_{\mathbf r'}\rangle,
\label{Bragg}\end{equation}
which is only nonzero below the critical temperature where
$\langle\sigma_{\mathbf r}\rangle\equiv{k'}^{1/4}$. This
$\mathcal{F}({\mathbf q})$ is proportional then to the Bragg delta-function
diffraction pattern that crystallographers would be most interested in.

\begin{figure}[htbp]
\begin{center}
\includegraphics[height=6.3cm]{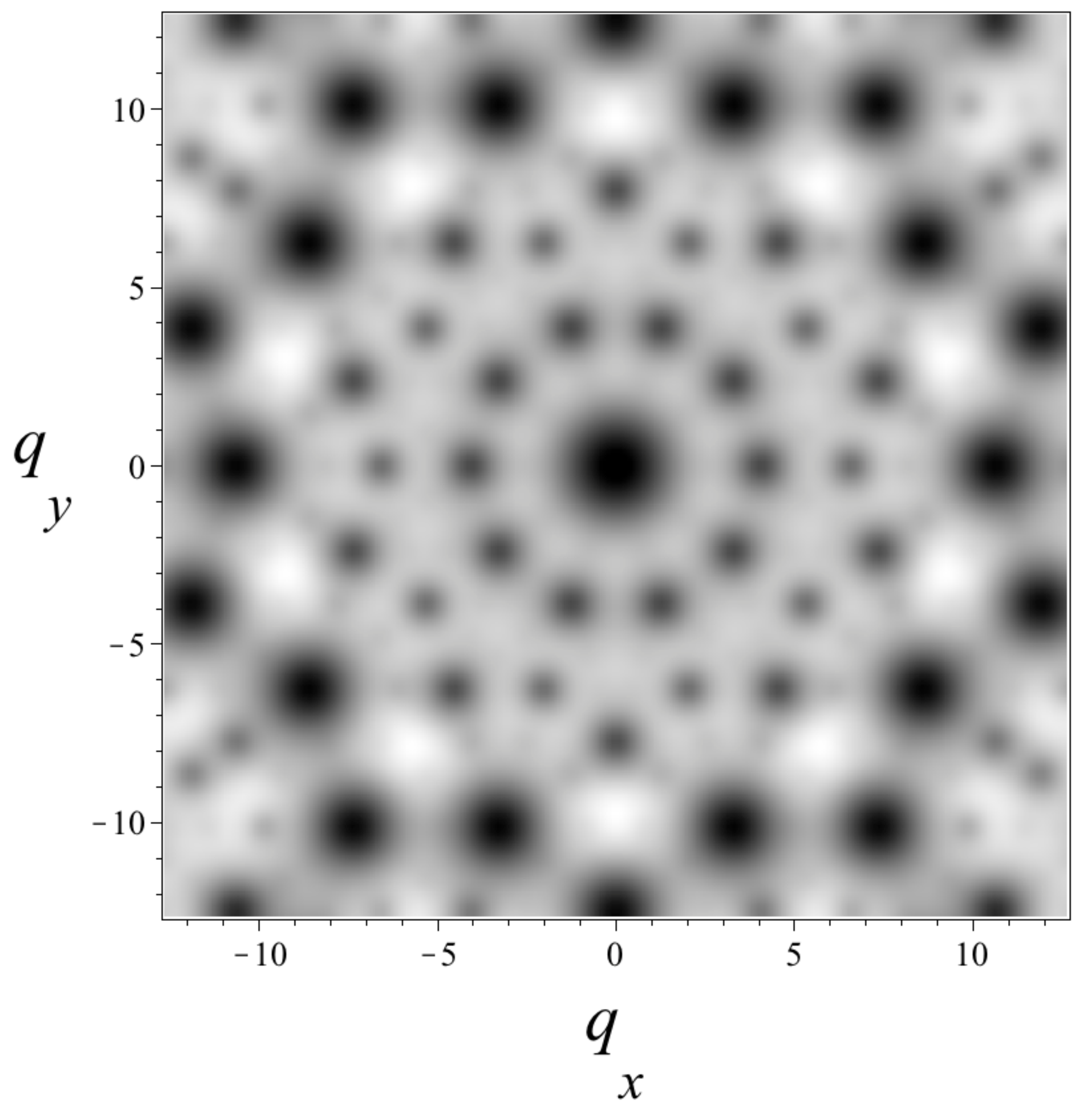}
\hspace{0.2cm}
\includegraphics[height=6.3cm]{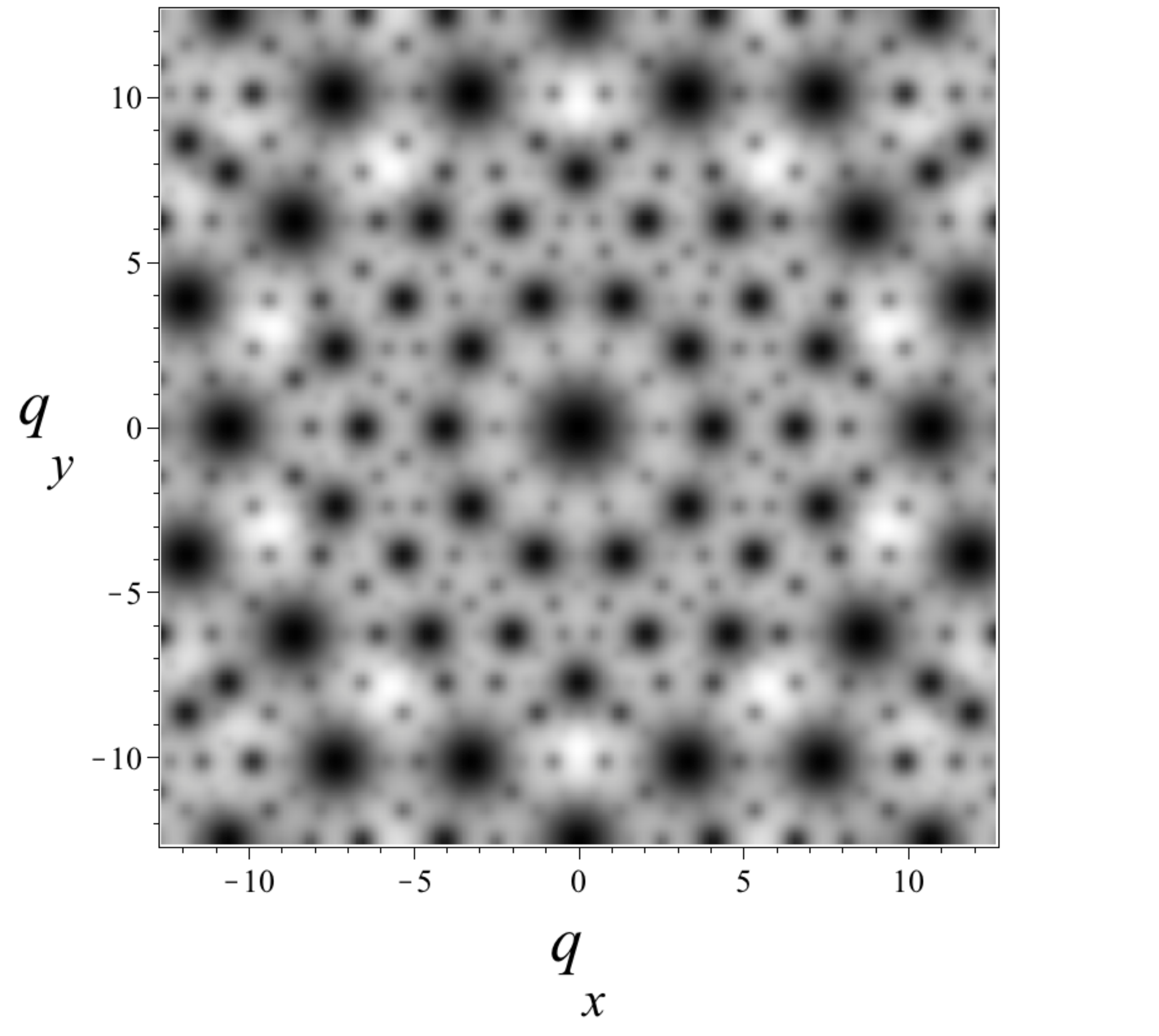}
\end{center}
\caption{Density plots of $\chi(q_x,q_y)$ for $k=0.7$. On the left:
low-temperature case; on the right: high-temperature case.}
\label{fig7}\end{figure}

The $\chi({\mathbf q})$ can be evaluated numerically to high
precision. To do this we sum first over contributions with
$\mathbf{r}-\mathbf{r}'$ about the same. This is defined by taking
two of the five grids of the pentagrid forming a periodic lattice
with rhombus faces and taking all contributions of spins in pairs of
such faces of this lattice that only differ by the same translation vector.
For such a partial sum there is only little variation in the number
of rapidity lines of each of the five grids passing between the spins.
Finally we have to sum over all different translation vectors that give
a significant contribution.

All these pieces were brought together in a long Maple program, enabling
us to evaluate $\chi({\mathbf q})$ for given $\mathbf q$ to high
precision \cite{APpenta}. As the elliptic modulus of the pentagrid
Ising model approaches one, we see more and more peaks forming. In
figure \ref{fig7} results are shown for $k=0.7$. Below the critical
temperature part of the intensity is in Bragg peaks (Dirac delta functions)
due to $\langle\sigma_i\rangle\equiv{k'}^{1/4}\ne0$, which are not
included in $\chi(q_x,q_y)$. There are no magnetic Bragg peaks above the
critical temperature.

This explains why the $\chi({\mathbf q})$ pattern is more intense at the
same $k$ value above the critical temperature. At the critical value
of the elliptic modulus,  $k=1$, the $\chi(q_x,q_y)$ obtains everywhere
dense power-law divergences, while remaining integrable over finite domains.
Below criticality, the Bragg delta-function peaks of (\ref{Bragg}) are also
everywhere dense, while $\mathcal{F}({\mathbf q})$ is similarly integrable
over finite domains.


\subsection{Some final remarks}

We note that we have also calculated $\chi({\mathbf q})$ for Ising
models on periodic lattices, but with Fibonacci-type variation
of the interactions \cite{APfibo}, which is a much easier problem.
Comparison with the pentagrid Ising model results allows us to see what
different effects variations of interactions or of lattice structure can have.

Finally, in this section we have shown how crucial de Bruijn's pentagrid construction
is for several of the steps in the calculations within the pentagrid Ising model
\cite {APpenta}, adding some new explanations not present in our earlier work.
The pentagrid picture of de Bruijn is truly powerful for calculations on
Penrose tilings.

\begin{figure}[htbp]
\begin{center}
\includegraphics[height=5cm]{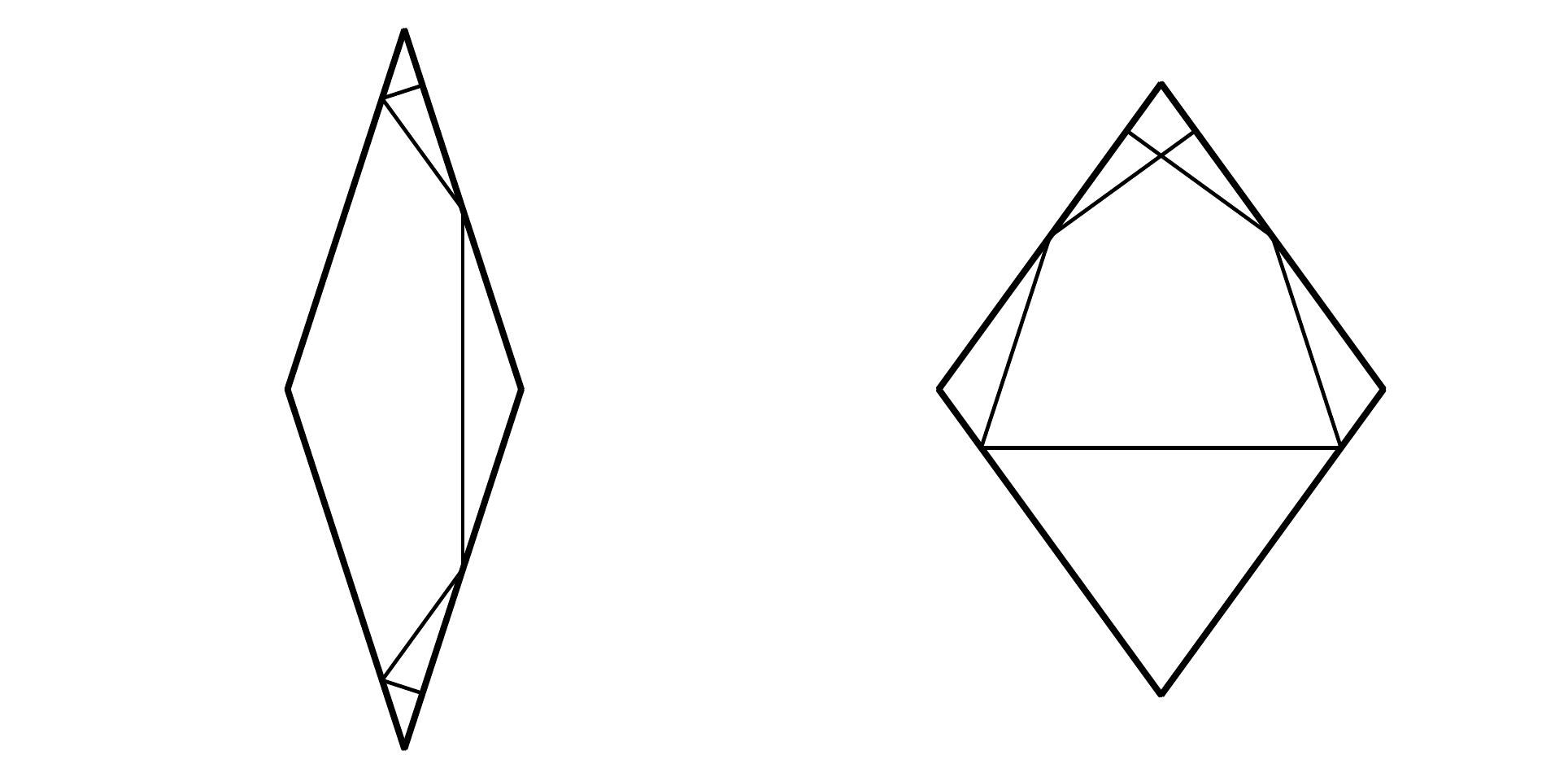}
\end{center}
\vspace*{-0.5cm}
\caption{The skinny and fat rhombi with Ammann lines.}
\label{fig8}\end{figure}

This has not always been properly understood, as a number of people have
given priority to Robert Ammann for his Ammann bars, ignoring de Bruijn's
work altogether. For the Penrose rhombus tiling one gets the Ammann bars by
decorating the rhombi with special Ammann lines as in figure \ref{fig8},
providing an alternative representation for Penrose's matching rules. Assuming
that the rhombi have sides of length 1, two sides of each rhombus are split into
pieces of length $\cos(\pi/5)$ and $1-\cos(\pi/5)$, whereas the other two sides
are divided in three with lengths $1/2$, $[1-\cos(2\pi/5)]/2$ and $\cos(2\pi/5)/2$,
compare also \cite{Stehling}.

If the rhombi are properly matched together these Ammann line pieces join
to form the straight lines that are now commonly called Ammann bars,
see \cite[fig.~7]{SoSt} or \cite[fig.~5]{Ste}. Unlike de Bruijn's pentagrid,
the five Ammann grids are not equidistant, as the spacings now follow a
Fibonacci pattern \cite{SoSt,Ste,Luck}.

At this moment it is not at all clear how the work of \cite{APpenta} could
be redone without significant extra complications using Ammann bars, even though
de Bruijn has argued \cite{Bruijn8}, that the Ammann quasigrid is topologically
equivalent to the pentagrid. In particular, many details of \cite{APpenta}
reviewed in subsection \ref{prob} depend on ratios of areas and simple
translations, exploiting the grids being equidistant.

In closing, note from figure \ref{fig8} that the Ammann bars intersect once on
each edge of the Penrose tiling. This tempts us to propose a new Yang--Baxter
integrable Ising model with Ising spins on all vertices and Ising interactions
along all edges, with two especially chosen but different interactions for edges
intersected by two or by three Ammann bars. Even though there seems to be no
difficulty to obtain the free energy per edge in the thermodynamic limit, the
determination of pair correlations cannot be done just following \cite{BaxZI},
as an odd number of rapidity lines can cross between spins, e.g.\ three for
a neighbor pair.


\section*{Acknowledgments}

This work was supported in part by the National Science Foundation
under grant No.\ PHY-07-58139. All figures were prepared with Maple 17.


\end{document}